\documentclass[twocol]{ametsocV6.1}

\usepackage{soul}
\usepackage{enumitem}

\title{Diurnal Warm Layer and Marginal Instability in the Ocean Surface Boundary Layer}

\authors{Adam Hall,\aff{a}
Hieu T. Pham,\aff{a}
Sophia Merrifield,\aff{b}
Sutanu Sarkar,\aff{a,b} \correspondingauthor{ssarkar@ucsd.edu}
}

\affiliation{\aff{a}{Mechanical and Aerospace Engineering, University of California San Diego, La Jolla, CA, USA}\\
	\aff{b}{Scripps Institution of Oceanography, University of California San Diego, La Jolla, CA, USA}
}

\abstract{Large-eddy simulations with high near-surface grid resolution are used to investigate the diurnal modulation of the ocean surface boundary layer (OSBL) at various latitudes for Coriolis frequencies corresponding to 15-70\textdegree N. The OSBL, which is driven here by a low wind, surface waves and a diurnally varying heat flux, exhibits a diurnal warm layer (DWL) during daytime and a convective mixing layer at night. The DWL exhibits enhanced shear leading to daytime turbulence that is found in this study to exist in a state of marginal instability (MI) over a wide range of latitudes and convective heat fluxes. The MI is characterized by an intricate balance between the momentum input by wind, stable stratification owing to solar radiation and turbulent mixing in the DWL, such that the gradient Richardson number is maintained around the critical value of 0.25 from morning to late afternoon. The enhanced shear and stratification sustained by MI leads to a substantial deviation from Monin-Obukhov Similarity Theory. Turbulent mixing induced by MI is shown to be similar to isolated local Kelvin-Helmholtz shear instability events and to share a similar range of flux coefficients and flux Richardson numbers.
The impacts of increasing rotation and nightly convection on OSBL variability at the inertial and diurnal timescales are quantified through a turbulent kinetic energy analysis. The interaction between inertial oscillations and diurnal shear is an important feature of the DWL at mid and high latitudes.}

\begin{document}

\maketitle

%
%
%
\statement
On sunny days with light winds, a thin layer of warm water forms in the ocean and flows as a surface jet. This layer is important to heat transfer between the ocean and atmosphere. We show that across latitudes ranging from the tropics to the Arctic, and under multiple heating conditions, inputs from the wind and sun are balanced in this layer by a state of turbulent mixing different from what was previously thought. This unique mixing alters temperature and velocity relative to predictions from simplified ocean models and changes the amount of heat absorbed by the deep ocean. These results reveal a critical gap in our understanding of turbulent mixing, highlighting the need for improved weather and climate models.

%
%
%

%

\section{Introduction}

The OSBL constantly evolves at multiple timescales due to the continuous input of momentum by wind and waves, buoyancy by surface heat fluxes and turbulent mixing. On days with weak to moderate winds and moderate to strong solar irradiation, the interplay between wind and stable solar flux leads to the formation of a diurnal warm layer (DWL) in the upper few meters of the ocean during daytime. The DWL can induce sea surface temperature (SST) anomaly up to a few degrees, and thus, have consequential impacts on the air-sea interaction and thermal transport from the surface to the ocean interior \citep{WijesekeraWJ:2020, HughesMS:2021, ZeidenTSD:2024, WitteZ:2025}.
A diurnal surface jet also develops alongside the DWL \citep{PriceWP:1986}, but its effects on turbulent mixing and heat transport remain poorly understood. At night, the OSBL is driven by surface cooling, and the transition between the strongly-stratified turbulence of the DWL to convective turbulence impacts vertical transport. In the present study, we use high-resolution LES to investigate the turbulent processes that modulate the OSBL at diurnal and inertial timescales.

While there have been numerous observational and numerical studies of the DWL at low latitudes in recent years \citep{HughesMS:2021, WitteZ:2025, SarkarP:2019}, less is known about the DWL at mid and high latitudes.
Recent field observational campaigns and LES studies have revealed characteristics that are unique to the high-latitude DWL \citep{MiraccaLageMS:2025, HauserTW:2026}. 
First, owing to the large value of $f = 2\Omega \sin\phi$, where $\Omega$ is the rotation rate of Earth and $\phi$ is latitude, the local diurnal jet vector at high latitudes rotates significantly faster than in the tropics, completing a rotation in $\sim 12\text{--}14$ hours rather than days. This rapid rotation of the shear vector continuously alters the alignment between the surface wind stress vector and near-surface currents. Consequently, high $f$ tends to suppress entrainment at the base of the DWL compared to the tropics, altering how heat and turbulent kinetic energy (TKE) are transferred downward into the thermocline. Second, during summer months, high-latitude regions experience continuous or near-continuous solar radiation. Under light-to-moderate wind conditions, this long-duration heating produces extreme diurnal SST anomalies, e.g., between 1.5 to 7.4$^\circ\text{C}$ in localized Arctic waters \citep{JiaML:2023, HauserTW:2026}. The role of turbulent mixing in the context of these high-latitude features is not well understood and is one focus of the present LES study. 

General ocean or climate models cannot resolve DWL turbulence and often rely on a 1-D mixing model as a turbulent mixing parameterization. Monin-Obukhov Similarity Theory (MOST) is often used for this purpose in the near-surface boundary layer. Based on dimensional analysis, MOST suggests that the shear and stratification weaken with distance from the surface. However, DWL observations indicate a local enhancement of shear and stratification with increasing depth in the OSBL \citep{KudryavtsevS:1990, SutherlandMR:2016, MoulinMS:2018, HughesMS:2020a}, putting the validity of MOST into question. Previous LES studies of OSBL with convective surface forcing have shown that Langmuir circulations can cause deviations from MOST \citep{LargePD:2019, ZhengHD:2021,GiddyWF:2026}. The effects of Langmuir turbulence on the OSBL with diurnal heat flux have been investigated at moderate-to-high winds \citep{NohGR:2009, KukulkaPS:2013, PearsonGPB:2015}. When the OSBL is strongly stabilized by solar radiation at low winds, Langmuir circulations are suppressed, and it is unclear whether MOST is valid in the strongly stratified regime of a DWL \citep{MinN:2004, PanL:2026}. The present LES, which include the effects of Langmuir turbulence, will address this question.

Observations of enhanced shear and stratification \citep{SutherlandMR:2016, MoulinMS:2018, HughesMS:2020b} motivate further study of how turbulence evolves in the DWL. 
Observations are often limited,
making it difficult to identify turbulence and its relation to the diurnally evolving background.
 Using idealized 1-D vertical simulations, \cite{HughesMS:2020a} suggested that DWL shear evolves differently when the wind is above or below 2 m s$^{-1}$. 
{Further, they found that} above this speed, the wind work is sufficient to overcome the stabilizing buoyancy flux and induce marginal instabilities (MI), agreeing with their observations. Marginal instabilities are a pathway to turbulence in which the interplay between elevated shear and stratification helps sustain turbulence over a long time period while the gradient Richardson number clusters around the critical value of 0.25 for shear instability \citep{ThorpeL:2009}. In a subsequent study, \cite{HughesMS:2021} performed 2D simulation of a DWL at a moderate wind speed 
and showed that the DWL may not exhibit marginal instabilities (MI) since their simulations show that, in the morning and early afternoon, the gradient Richardson number is smaller than 0.25 but, nevertheless, instabilities are small. In the present study, we use 3D LES with high grid resolution and Langmuir turbulence to revisit whether MI exists in a DWL with a moderate wind speed of $\sim$ 3 m s$^{-1}$. The establishment of MI is necessary to guide future improvement of 1-D mixing models such as those in General Turbulence Ocean Models \citep{LiBB:2021}. \cite{JohnsonFL:2023} have demonstrated that, when forced with strong diurnal warming, these models exhibit biases relative to LES results.

The manuscript is organized as follows. The numerical method and the model setup are provided in section~\ref{sec:model}. The overall diurnal and inertial variability of the boundary layer are described in section~\ref{sec:ML}. In section~\ref{sec:MI}, we address whether MI exists in the DWL, examine the validity of MOST, and diagnose the turbulent kinetic energy (TKE) balance. The impacts of changing latitudes and surface cooling flux on DWL characteristics are explored in sections~\ref{sec:lat} and \ref{sec:convection}, respectively. We provide a discussion of the present study in section~\ref{sec:conclusion}.

\section{Numerical Setup} \label{sec:model}

The LES solves the wave-averaged Boussinesq equations for the grid-filtered Eulerian velocity components ($u_i$), and temperature ($T$). The governing equations are as follows:
\begin{equation*}
	\begin{gathered}
	\begin{split}
		\frac{\partial u_i}{\partial x_i} = &0, \\
		\frac{Du_i}{Dt}= & \epsilon_{ijk}f_k\left(u_j+u_j^s\right)+\epsilon_{ijk}\omega_ku_j^s-\frac{\partial\pi}{\partial x_i} \\
		& +\alpha_T\left(T-T_0\right)g\delta_{i3} +\nu\frac{\partial^2u_i}{\partial x_j^2} -\frac{\partial\tau_{ij}^{sgs}}{\partial x_j},\\
		\frac{DT}{Dt}= & -u_j^s\frac{\partial T}{\partial x_j}+\kappa_T\frac{\partial^2  T} {\partial x_j^2}\\
		& -\frac{1}{\rho_0c_p}\left(\frac{\partial Q_j^{T,sgs}}{\partial x_j}-\frac{\partial Q_{sw}}{\partial x_3}\right),
	\end{split}
	\end{gathered}
\end{equation*}
where $\frac{D}{Dt}=\frac{\partial}{\partial t}+u_j\frac{\partial}{\partial x_j}$ is the material derivative. The generalized pressure, $\pi$, is defined as
\begin{equation*}
\pi=\frac{p}{\rho_0}+\frac{2e}{3}+\frac{1}{2}\left[\left|u_i+u_i^s\right|^2-\left|u_i\right|^2\right] \, , 
\end{equation*}
with $p$ denoting the dynamic pressure and $e=\frac{1}{2\tau^{sgs}_{ii}}$ the subgrid turbulent kinetic energy. The Boussinesq term is computed using the coefficient of thermal expansion $(\alpha_T = 2.5 \times 10^{-4}$ K$^{-1})$. The linear equation of state, $\rho/\rho_0 = 1-\alpha_T (T-T_0)$, is used to compute density with the reference density $\rho_0 =1027$ kg m$^{-3}$. 

 \begin{figure*}[h!]
	\noindent\includegraphics[width=39pc,angle=0] {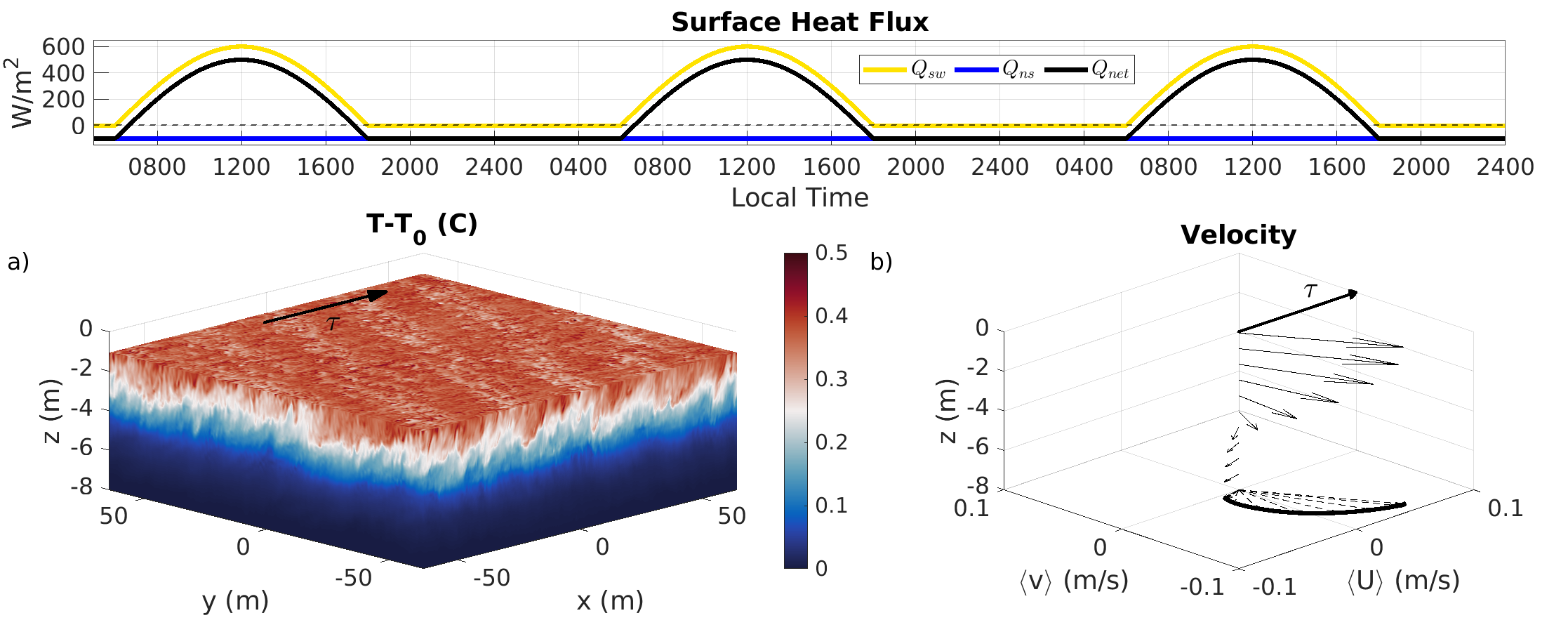}\\
	\caption{Depiction of coherent flow structures in the mixing layer at 1300 hrs in case L45H6C1: (a) 3D view of the temperature field; and (b) the Ekman spiral constructed from the horizontally-averaged zonal and meridional velocities. The applied diurnal solar heat flux ($Q_{sw}$), non-solar heat flux ($Q_{ns}$) and net flux ($Q_{net}$) are shown in the top panel. Note that the top surface in panel (a) is at 1-m depth. Black arrow in (a) indicates wind direction.}
	\label{fig:DWL_vol}
\end{figure*}

A constant wind stress, $\tau_w = 0.02 \, {\rm N}\, {\rm m}^{-2}$ (corresponding to a moderate wind speed $U_{10}$ of $3.1$ m s$^{-1}$) is applied in the zonal direction. The drag coefficient-based relationship of \cite{LargeP:1981} with $C_d = 1.7 \times 10^{-3}$ is used to correlate $U_{10}$ and $\tau_w$. The Stokes drift profile, $u^s(z)$, is obtained using the empirical theory wave model of \cite{LiFK:2017, LiBB:2021}.
The model assumes a wave spectrum of \cite{Phillips:1958} to obtain the surface Stokes velocity, $u^s_0 \approx 0.016 U_{10} = 4.96 \, {\rm cm} \, {\rm s}^{-1}$ and the peak wave number, $k_p= 1.13 \, {\rm m}^{-1}$.
A penetrative solar flux, $Q_{sw}(z,t)$, is applied via a 9-term exponential transmissivity model from \cite{PaulsonS:1981} assuming a Jerlov-type I water-type. 
The time-dependent component of $Q_{sw}$ at the surface is implemented using a sinusoidal function:
\begin{equation}
	Q_{sw}(t, z = 0) =\left\{
	\begin{array}{ll}
		Q_{sw,max} sin\left[\frac{2\pi}{T_d}(t - t_{0600})\right] \hspace{12pt} \\
		\hspace{44pt} 0600 \le t \le 1800 \, \text{hrs}, \\
		0 \\
		\hspace{44pt} 1800 \le t \le 0600 \, \text{hrs}, \\
	\end{array}
	\right.
	\label{eq:Qsw}
\end{equation}
where t is the local time. For each of the three simulated days, the solar heat flux begins at 0600 hrs (i.e., $t_{0600}$) of every day, peaks at the maximum value of $Q_{sw,max} = 600$ W m$^{-2}$ at 1200 hrs, and subsides at 1800 hrs, yielding a diurnal time period of $T_d$ = 24 hrs as shown in top panel of figure \ref{fig:DWL_vol}. Positive (negative) \emph{surface} flux indicates heat into (out of) the ocean.

We simulated six cases spanning a range of latitudes and cooling flux, $Q_{ns}$. Four cases with the same value of diurnal heat flux are performed at 15$^\circ$N, 30$^\circ$N, 45$^\circ$N, and 70$^\circ$N latitude, respectively. The diurnal heat flux in these simulations consists of a constant cooling flux, $Q_{ns} = -100$ W m$^{-2}$, and the time-dependent $Q_{sw}$ given by Eq.~(\ref{eq:Qsw}). 
Two additional cases are simulated at 45$^\circ$N with $Q_{ns} = 0$ and -200 W m$^{-2}$ to explore the impact of changing $Q_{ns}$ and thereby the net heat flux, $Q_{net} = Q_{sw}(t, z=0) + Q_{ns}(t)$.
The surface forcing leads to the important dimensionless control parameters of turbulent Langmuir number $La_t = (u_\ast/u_s)^{1/2}$, and Hoenikker number $Ho = 2B_0/(k_p u_0^s u_\ast^2)$ where 
 $B_0 = \alpha g Q_{net} / \rho_0 c_p$ 
 is the net surface buoyancy flux.
In the present study, the ratio of the frictional velocity to the surface Stokes drift velocity is given by $La_t = 0.3$, indicating that Langmuir turbulence, contributes significantly to near-surface mixing \citep{ McWilliamsSM:1997}. The Hoenikker number, which compares the surface buoyancy flux to the Stokes production, varies between -0.22 at night time and 0.56 at noon time (when $Q_{net}$ = 500 W m$^{-2}$). Note that positive (negative) $Ho$ denotes stabilizing (destabilizing) heat flux. 
In previous LES studies using monochromatic wave fields \citep{MinN:2004, PanL:2026}, Langmuir circulations were shown to break down for stabilizing buoyancy flux when the values of $Ho$ reach the range between 1 and 2. 

To ease discussion, we use the following abbreviations: L15, L30, L45 and L70 to denote the various latitudes; and C0, C1 and C2 to denote the $Q_{ns}$ values of 0, -100 and -200 W m$^{-2}$, respectively. 
 We use the naming convention LxHyCz to label the cases where x, y, and z correspond to the values of the latitude, $ Q_{sw,max}$ and $\vert Q_{ns}\vert$ in hundreds of W m$^{-2}$, respectively. For example, L70H6C1 denotes the case at 70$^\circ$N latitude, $ Q_{sw,max} = 600$ W m$^{-2}$ and $Q_{ns} = -100$ W m$^{-2}$.
All simulations are spun up with constant wind stress for 3 inertial periods for the Ekman layer to fully develop. After the diurnal heat flux is applied, they are continued for a span of approximately three calendar days with a time step as small as 0.5 s using a constant CFL number of 1.2. Each case takes approximately 45,000 computational hours distributed over 512 cores.

The computational domain is a rectangular box with $512 \times 512 \times 256$ grid points in the zonal (x), meridional (y), and vertical (z) directions in all cases except for the case at $15^\circ$N in which a larger domain with $N_x = N_y = 768$ is necessary to accommodate a thicker DWL. The grid is uniform in both horizontal directions with $\Delta x = \Delta y = 0.25$ m. The vertical grid spacing is 0.05 m at the surface, stretching at a rate of 3\% until a spacing of 0.25 m is reached at z = -9.25 m. 
The fine vertical grid resolution near the surface ensures that the LES resolves the Monin-Obukhov length scale, $L = u_\ast^3/\kappa B_0$, where the von Karman constant $\kappa = 0.4$. In the case with the strongest heat flux, $Q_{net} = 600 \, W \, m^{-2}$, $\Delta x = \Delta y = 0.4 L$, and $\Delta z = 0.08 L$ at the surface. 
The grid is uniform with $\Delta x = \Delta y = \Delta z = 0.25$ m between 9.25 and 40 m, below which it stretches at a rate of 2.5\%. A sponge layer is applied between 59 and 72 m wherein temperature and velocity fluctuations are damped to zero using a Rayleigh damping function.
The numerical methods used in the LES model are similar to those in \cite{PhamS:2018}. The LES uses a 2$^{\text{nd}}$-order central difference in space and a time-advancement scheme with mixed explicit 3$^{\text{rd}}$-order Runge-Kutta and 2$^{\text{nd}}$-order Crank-Nicholson methods. Generalized pressure is calculated with a multigrid Poisson solver. The filtered structure function to compute subgrid stresses $\tau^{sgs}_{ij}$, follows the model of \cite{DucrosCL:1996}, as further refined for a shear layer by \cite{VanDinePS:2020}. A subgrid Prandtl number of one is used to calculate the subgrid heat flux, $Q^{T,sgs}_j$. 

Several turbulence statistics will be reported as Reynolds averages, which are computed here using averaging over a horizontal plane. The turbulent kinetic energy is given by TKE = $(1/2) \langle u'_i u'_i \rangle$, where the angle brackets denote Reynolds average. The TKE production from the along-wind shear, $P_u$, and cross-wind shear, $P_v$, is given by $P= P_u + P_v = -\langle u'w'\rangle \partial \langle u + u^s \rangle /\partial z - \langle v'w'\rangle \partial \langle v\rangle /\partial z$.
The turbulent dissipation is given by $\varepsilon=-\langle \tau_{ij}'s_{ij}'\rangle+2\nu\langle s_{ij}'s_{ij}'\rangle$ and includes both the resolved and subgrid contributions. Here, $\tau_{ij}'$ and $s_{ij}'$ are the fluctuating subgrid stresses and resolved strain. The buoyancy flux is given by $B=\frac{-g}{\rho_0}\langle \rho'w'\rangle$, and vertical transport $dT_3/dz$ is computed using $T_3 = 0.5 \left<u_i'u_i'u_3'\right> + \frac{\left<u_3'p'\right>}{\rho_0} -  2\nu\left<u_i's_{i3}'\right> + T_{3,sgs}$.

\section{Characteristics of the diurnal mixing layer} \label{sec:ML}

 In this section, we describe the results of case L45H6C1 to demonstrate how the wind stress, surface wave forcing, diurnal surface heat fluxes, and inertial oscillations modulate the temperature and velocity fields within the surface boundary layer. Over the simulated three-day period, this structural modulation at mid- and high latitudes is driven by near-surface Langmuir turbulence, alongside vertical entrainment from both shear and convective turbulence at the base of the mixing layer. Notably, despite the strong stabilizing solar heat flux that peaks at noon, shear-driven entrainment remains significant during the day. 
 This entrainment is concentrated within a marginal instability (MI) layer where the local gradient Richardson number ($Ri_g$) clusters around the critical threshold of 0.25 for shear instabilities. 

Figure~\ref{fig:DWL_vol}(a) illustrates the temperature field in the OSBL at 1300 hrs on the first day, when $Q_{net} = 480$ W m$^{-2}$. 
On the horizontal plane at $z = -1 \, m $, shear instabilities organize into coherent rows. The structures at this time are \emph{not} signature of Langmuir circulations.
The surface current at this time is oriented approximately $58^\circ$ to the right of the wind direction. 
At the ML base, vigorous turbulent entrainment is driven by shear instabilities, as visualized in the vertical planes of Fig.~\ref{fig:DWL_vol}(a). The orientation of these deeper shear instabilities is dictated by the local shear vectors at depth. At the Equator ($f = 0$), the directions of the wind and current shear in the ML are closely aligned. Away from the Equator, however, planetary rotation rate ($f > 0$) induces Ekman veering, which complicates the ML structure and reduces the entrainment rate, ultimately reducing thermal transport across the layer \citep{PhamS:2017}. 

 \begin{figure*}[t!]
	\noindent\includegraphics[width=39pc,angle=0] {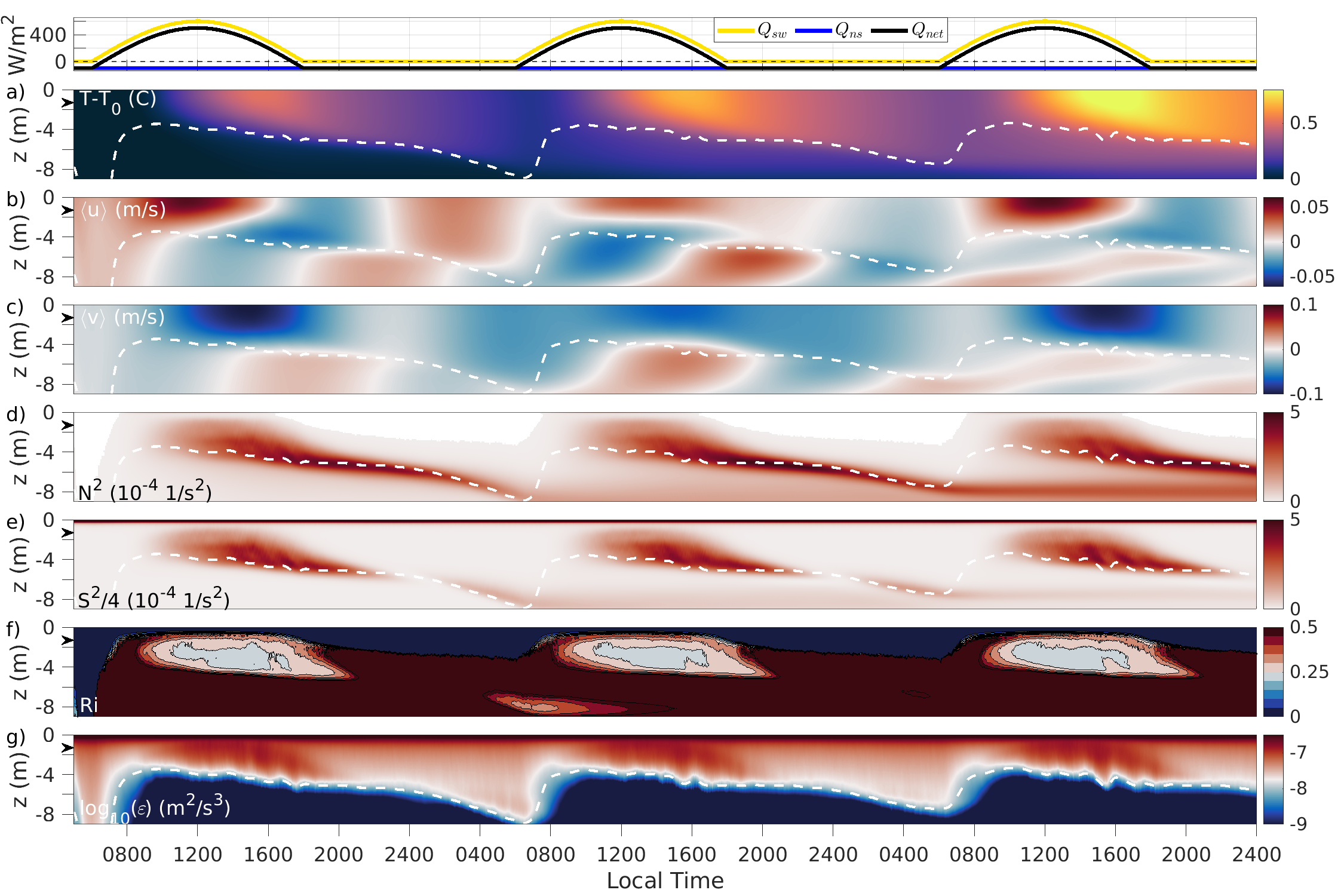}\\
	\caption{ Modulation of the mixing layer by diurnal heat fluxes, inertial oscillations and turbulence in case L45H6C1: (a) temperature deviation, $T - T_0$; (b, c) zonal and meridional velocity components, $\left<u\right>$ and $\left<v\right>$; (d) stratification, $N^2$; (e) reduced Lagrangian squared shear, $S^2/4$; (f) gradient Richardson number, $Ri_g$; and (g) dissipation rate, $\log_{10} (\varepsilon)$. Top panel shows the surface heat fluxes. Arrows on the vertical axes mark the base of the wave-induced Stokes layer. Dashed white lines in this figure and hereafter denote the mixing layer depth above which $\varepsilon > 10^{-8}$ m${^2}$ s$^{-3}$.}
	\label{fig:DWL_background}
 \end{figure*}

The flow field shown in Fig.~\ref{fig:DWL_vol} varies significantly in time due to the variability in surface heat flux and the spontaneous development of inertial oscillations (IOs) as depicted in Fig.~\ref{fig:DWL_background}. 
The horizontally-averaged temperature and the mixing layer depth (MLD) in Fig.~\ref{fig:DWL_background}(a) indicates that, typically, the absorption of solar radiation leads to daytime warming in the top 4-m diurnal warm layer (DWL), and nocturnal cooling in the convective mixing layer (CML) of approximately 8-9 m.
Here, the MLD is defined as the greatest depth from the surface where the dissipation rate ($\varepsilon$) exceeds $10^{-8}$ m$^2$ s$^{-3}$. {This is a turbulent mixing layer depth, distinct from a temperature- or density-defined mixed-layer depth}.
We note that the initial rapid ML deepening during the first two hours is the transient response to the sudden application of surface cooling flux at the beginning of the simulation, which amplifies the IOs. The IOs that develop during the spin-up persist throughout the 3-day period in all simulations.
On a daily basis, the shallowest MLD (3.4 m) occurs at approximately 1000 hrs, and it is deepest (9.1 m) at 0700 hrs after $Q_{net}$ changes sign. 
The deepening rate of MLD between these two times vary significantly. 
Notably, the MLD increases slightly due to turbulent entrainment even as the solar radiation (a stabilizing influence) grows to its peak value between 1000 hrs and noon. The sea surface temperature (SST) reaches its daily peak value at approximately 1400 hrs. During the afternoon and until midnight, the MLD grows at a faster rate. The MLD grows fastest between midnight and 0700 hrs of the following day during the period that convective turbulence dominates in the ML. The variation of SST and sea surface speed (SSU) will be quantified by line plots in a later section. 

The DWL and IOs modulate the Ekman current as illustrated by Fig.~\ref{fig:DWL_background}(b, c). 
On a daily basis, the near-surface zonal velocity ($\left<u\right>$) reaches its peak positive value at approximately noon, two hours after the MLD shoals to its shallowest value. The near-surface peak of negative meridional velocity ($\left<v\right>$) occurs approximately two hours after peak $\left<u\right>$. 
Due to the Stokes-enhanced Ekman veering, the cross-wind velocity component becomes larger than the along-wind component \citep{McWilliamsSM:1997}.
The surface speed, which strengthens during the day owing to the trapping of wind-driven momentum in the DWL, weakens during the night due to convective mixing of momentum. 
Below the MLD, the velocity components also exhibit oscillatory motions but at the inertial time period of 16.9 hours, noting that the bands of positive / negative velocity components are shorter than the 24-hour bands of positive / negative velocities above the MLD. 
The positive near-surface $\left<u\right>$ between 2200 hrs and 0700 hrs during the first night also reflects the presence of IO. 
The diurnal and inertial oscillations can interact with each other. Over the three-day period, the surface speed at noon on the first and third days is notably stronger relative to the second day.
When the diurnal and inertial oscillations act in phase, the Ekman velocity in the DWL is enhanced resulting in stronger shear and turbulent mixing. 

The evolution of stratification ($N^2$) and reduced {Lagrangian} squared shear ($S^2/4$) in the ML exhibits strong diurnal modulation with significantly larger values during the day (Fig.~\ref{fig:DWL_background}d, e). 
As the DWL develops, the solar radiation stabilizes the ML (i.e, increasing $N^2$), and the Ekman current accelerates leading to a stronger surface jet (i.e, increasing $S^2$). 
 Between 0800 to 1800 hrs, the vertical profiles of $N^2$ and $S^2$ show peak values in the middle of the ML, not at the base of the Stokes layer or at the MLD location. 
 In contrast, the nightly variation between 1800 and 0600 hrs exhibits thin layers of enhanced $N^2$ and $S^2$ with peak values coinciding with the MLD. 
 
  \begin{figure*}[t!]
 	\noindent\includegraphics[width=39pc,angle=0] {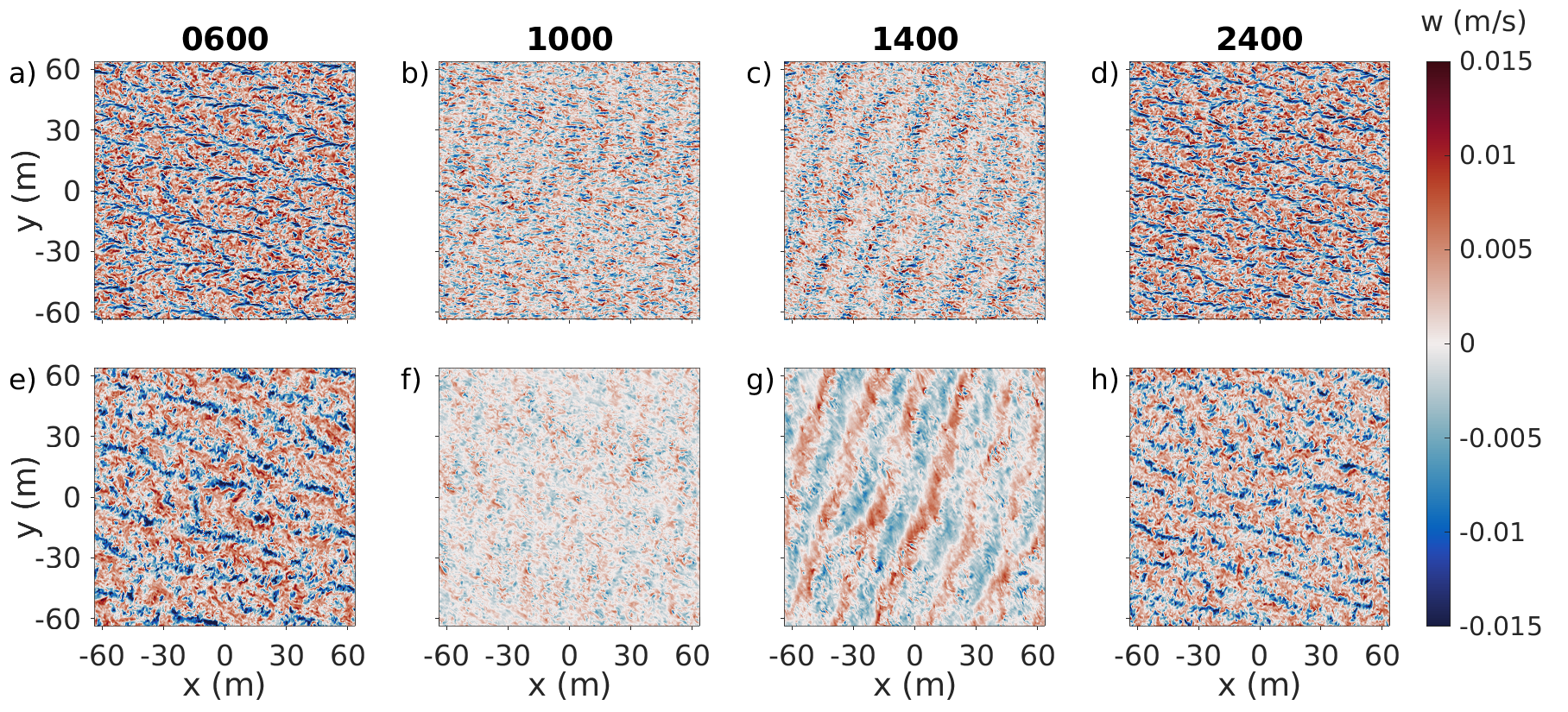}\\
 	\caption{Horizontal slices of vertical velocity ($w$) at 1-m (top row) and 3-m (bottom row) depths in case L45H6C1 illustrate how turbulence structures evolve as the ML transitions from convective mixing to stratified marginal instabilities. Four times are shown during the second day of the simulation: (a, e) 0600 hrs (b, f) 1000 hrs; (c, g) 1400 hrs; and (d, h) 2400 hrs.}
 	\label{fig:MI_w}
 \end{figure*}
 
 The difference in the evolution of $N^2$ and $S^2$ profiles between day and night suggests turbulence manifests differently as the state changes between the two opposite ends of mixing regimes - stratified turbulence in the DWL and convective turbulence in the CML. As a result, different layers form, each with its own distinctive turbulence characteristics, as summarized below: 

\begin{enumerate}[label=(\roman*)]
\item Stokes layer: A layer at the surface (1.3 m thick for this study) that is defined by the Stokes velocity profile that remains invariant among cases. It has the strongest (Lagrangian) shear and turbulence over the entire water column.

\item Mixing layer (ML): A diurnally-varying layer whose lower boundary is defined by a threshold of $\varepsilon > 10^{-8}$ m$^2$ s$^{-3}$. It includes and extends beyond the Stokes layer, and contains turbulence driven by Eulerian shear and/or convection.

\item Diurnal warm layer (DWL): This layer refers specifically to the day-time ML from 0600 to 1800 hrs. It is characterized by stratified turbulence that transports heat from the surface into the thermocline.

\item Convective mixing layer (CML): This layer refers to the night-time ML. There is upward heat transport by convective turbulence in this layer and $N^2$ takes negative values (bright white in Fig.~\ref{fig:DWL_background}(d)). 

\item Descending shear layer / transition layer (TL): A thin layer with enhanced $N^2$ and $S^2$ at the base of the ML as the DWL evolves into the nightly CML as shown in Fig.~\ref{fig:DWL_background}(d,e). The enhanced shear in this layer drives turbulent entrainment of cold fluid from the thermocline into the ML. Different from the CML, the TL is characterized by stable stratification ($N^2 > 0$). Note that the nightly ML consists of the CML above the TL, with the turbulent heat flux flipping sign between these two layers.
\end{enumerate}

Of the layers mentioned above, the DWL uniquely exhibits stratified turbulence as indicated by the gradient Richardson number ($Ri_g = N^2 / S^2$ where $S^2$ here is the Lagrangian squared shear) and turbulent dissipation rate ($\varepsilon$) in Fig.~\ref{fig:DWL_background} (f, g). Despite the stabilizing solar radiation, the strongest $\varepsilon$ occurs in the DWL reaching at least an order of magnitude larger peak than that found in the CML or TL.
Turbulence is driven by marginal instabilities (MI) characterized by strong local Lagrangian $S^2$ and $N^2$ with the corresponding $Ri_g$ values clustering around the critical value of 0.25. 
Different from textbook shear instabilities that develop linearly from normal modes, MI involves highly nonlinear physics that sustains turbulence through an intricate balance between $N^2$ and $S^2$. While turbulent mixing tends to reduce local $N^2$ and $S^2$, the momentum input by wind increases $S^2$ and the solar radiation increases $N^2$. Of interest is the interplay between wind, solar heat flux and turbulence that leads to the state of MI in the DWL, motivating our further examination of MI physics in the next section.

Horizontal planes of vertical velocity ($w$) in Fig.~\ref{fig:MI_w} depict the MI-induced structural modification of vertical motion as the ML transitions from night-time convection to daytime stratified shear flow. 
The figure shows snapshots at 1-m depth (i.e., the base of the Stokes layer) and at 3-m depth (i.e, the base of the DWL at 1000 hrs). 
Early in the morning at 0600 hrs when $Q_{net} < 0$, the ML exhibits multiple thin rows of negative $w$ with a horizontal spacing of approximately 25 m at both depths (Fig.~\ref{fig:MI_w}a,e). 
These rows are induced by counter-rotating Langmuir circulations, and are mostly oriented along the direction of wind with an offset to the right of approximately $15$\textdegree.
 The rows are not always parallel noting multiple {Y-junctions where adjacent rows intersect}. 
The wavelength of these cells is approximately three times the MLD. The orientation and the aspect ratio of the cells agree with the observations in \cite{Faller:1964} and \cite{SmithPW:1987}.

As the solar radiation warms the ML during morning, a stable stratification develops, leading to significantly suppressed $w$ at 1000 hrs at both 1-m and 3-m depth (Fig.~\ref{fig:MI_w}b, f). 
The coherent rows of negative $w$ seen at the earlier time of 0600 hrs due to Langmuir circulations are no longer visible and the turbulence structures have smaller horizontal scales. The value of $Ho = 0.53$ at this time does not reach the range between 1 and 2 for the Langmuir circulations to be completely suppressed \citep{MinN:2004, PanL:2026}; however, the circulation-driven velocities are significantly weaker relative to the contribution of the shear instabilities developing at 3-m depth. It is further noted the $Ho$ criteria for the suppression of Langmuir circulations was computed using a monochromatic wave field in the previous studies while we use the peak wavenumber in the equilibrium broadband spectrum to compute $Ho$ in the present study.
While the $w$ field has weakened from 0600 to 1000 hrs, both $S^2$ and $N^2$ have increased (recall Fig.~\ref{fig:DWL_background}d, e). 

At 1400 hrs (Fig.~\ref{fig:MI_w}c, g), $w$ is organized into coherent tilted bands at both 1-m and 3-m depths, and they resemble the phase lines of quasi-2D Kelvin-Helmholtz (KH) rollers \citep{VanDinePS:2021}. 
Unlike the convective wind rows seen earlier at 0600 hrs, these bands are more coherent and have wider horizontal extent. They are advected together in parallel (i.e., they do not intersect). The bands at 3-m depth are relatively more coherent than those at 1-m depth despite having the same horizontal spacing. The horizontal wavenumber spectra of the w-field indicates a peak at the wavelength of $\lambda$ = 20.3 m, which is approximately 7 times the thickness of the shallowest MLD, another resemblance to KH shear instabilities \citep{SmythM:2000a, PhamSB:2009, MashayekCP:2013}. 
The bands propagate at the angle of $18.4^\circ$ to the right of the zonal direction.
Although the horizontal wavelength of shear instabilities at 1400 hrs is near that of the Langmuir cells seen at 0600 hrs, these instabilities result from different processes. The spacing of Langmuir circulation at 0600 hrs scales with the \emph{thick} convective nightly mixed layer such that the aspect ratio of the cells ranges between 1 and 3. In contrast, the wavelength of the shear instabilities at 1400 hrs scales with the \emph{thin} daytime mixed layer depth.

\section{Stratified shear instabilities in the DWL} \label{sec:MI}

We find that the DWL, despite the stabilizing surface heat flux and calm conditions with a moderate wind speed of $U_{10} = 3.1 $ m s$^{-1}$, supports shear instabilities and turbulence. These instabilities occur over a wide range of latitudes (from the Equator to the polar oceans) and $Q_{net}$ (from 0 to $600 {\rm W} \, {\rm m}^{-2}$). Notably, the simulations reveal that the DWL is {\em in a state of marginal instability (MI) in all cases considered here}. In this section, we use the results of case L45H6C1 to illustrate how DWL turbulence differs from Langmuir and convective turbulence. Since these shear instabilities develop close to the ocean surface, it is also of interest to explore how they cause the near-surface layer to deviate from the law of the wall that is integral to the Monin-Obukhov Similarity Theory (MOST).

\subsection{Coherent structures of shear-driven turbulence}

 \begin{figure*}[t!]
	\noindent\includegraphics[width=39pc,angle=0] {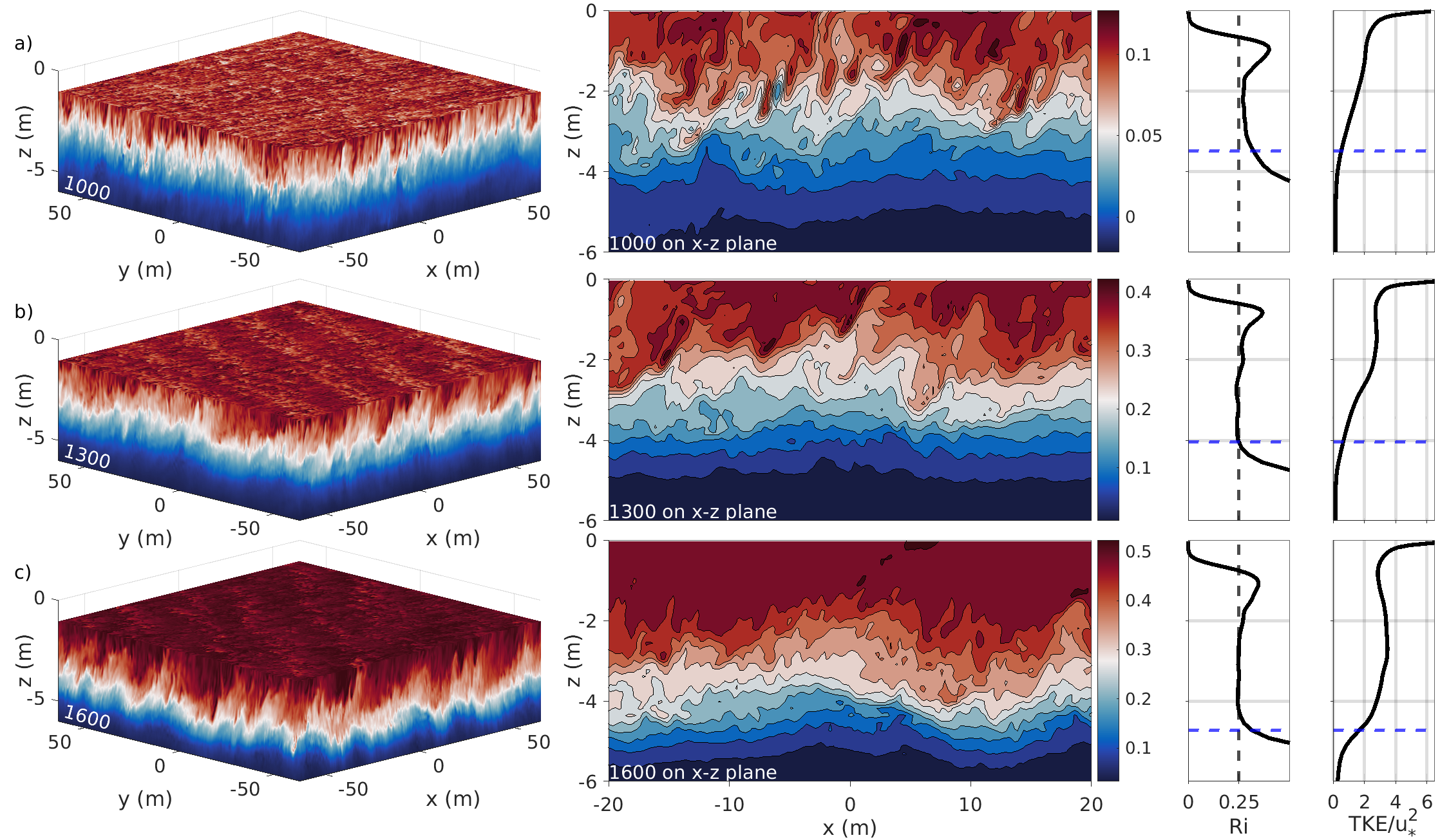}\\
	\caption{ The 3D fields (first column) and vertical x-z slices (second column) of temperature deviation depict the evolution of ML turbulence in case L45H6C1 at various times: (a) 1000 hrs; (b) 1300 hrs; and (c) 1600 hrs. The vertical profiles of $Ri_g$ (third column) indicate the presence of marginal instability with $Ri_g \approx 0.25$. The vertical TKE profiles (fourth column) show that turbulence develops a local peak of TKE inside the ML. Blue horizontal dashed lines mark the MLD while black vertical dashed lines mark the critical value of $Ri_g = 0.25$.}
	\label{fig:MI_tem}
 \end{figure*}

Three-dimensional views and vertical temperature slices in Fig.~\ref{fig:MI_tem}(a-c) illustrate downward heat transport from the surface. 
At 1000 hrs, energetic turbulence induced mainly by Stokes shear occupies $z > -3.5$ m, the region above the MLD. Beneath the MLD, turbulence becomes sporadic. The $Ri_g$ profile indicates a state of MI with $Ri_g$ values slightly larger than 0.25 below 1.5-m depth (Fig.~\ref{fig:MI_tem}a). As the ML deepens to 4-m depth at 1300 hrs, energetic fluctuations are seen with multiple large overturns of isotherms between 2-m and 4-m depth. The overturns in this region are a result of MI with $Ri_g$ dipping slightly below its critical value of 0.25. 
When the ML further deepens to 4.8-m depth at 1600 hrs, wavy isotherms with a wavelength of approximately 28.7 m are seen. 
TKE is enhanced as the $Ri_g$ values continue to cluster narrowly at the critical value. 
Different from the typical KH shear instability wherein the wavy normal-mode instability develops first and then evolves into billows that break down into subsequent turbulence, the wavy MI structures seen at 1600 hrs emerge after the occurrence of large overturns of isotherms at 1300 hrs. 
In isolated KH-driven mixing, the mean kinetic energy reservoir is depleted over time. In contrast, in the MI mixing scenario described here, momentum input by wind and heat input by solar radiation feed continuously into the mean kinetic energy reservoir by enhancing the local shear rate. The intricate balance between energy dissipation by turbulence and energy input by the wind helps maintain the MI state, thereby sustaining turbulence over a longer time period than during an isolated KH-billow event.

\subsection{Deviation from Monin-Obukhov Similarity Theory}

To further understand the consequences of DWL turbulence, we compare the LES results with the scaling laws of MOST \citep{MoninO:1954, Wyngaard:2010}. 
According to this theoretical framework, ML turbulence is characterized by the Monin-Obukhov (MO) length, $L = u_\ast^3/\kappa B_0$. 
Under neutral stability ($Q_{net} = B_0 = 0, L \rightarrow \infty$), turbulence is driven primarily by wind shear. In unstable conditions ($Q_{net} < 0, B_0 < 0, L < 0$), the ML is dominated by convection. In stable conditions as in the DWL ($Q_{net} > 0, B_0 > 0, L > 0$), solar radiation suppresses turbulence leading to reduced vertical transport. We note that, for the atmospheric boundary layer (ABL), the MO length is defined as $L = -u_\ast^3/\kappa B_0$ where $B_0$ is the surface buoyancy flux, again leading to a positive $L$ for the stable case when the ABL has $B_0 <0$.

Based on MOST, the vertical {Lagrangian} shear, $S = d\left<U\right>/dz = [ (\partial \left<u + u^s \right> / \partial z)^2 + (\partial \left<v\right> / \partial z)^2 ]^{1/2}$, the temperature gradient (\,$\partial \left< T\right> /dz$\,) 
and TKE production (\,$P=-\langle u'w'\rangle \partial \langle u + u^s \rangle /\partial z - \langle v'w'\rangle \partial \langle v\rangle /\partial z$\,) are modified as the stabilizing heat flux increases. Here, the angle brackets denotes horizontal average.
The modification is prescribed through the dimensionless parameter $\zeta = \vert z \vert /L$ and the universal stability functions for momentum, $\phi_m(\zeta)$, and heat, $\phi_h(\zeta)$, as follows:
\begin{eqnarray}
\frac{\partial \left<U\right>}{\partial z} &=& \frac{u_\ast}{\kappa \vert z \vert} \phi_m(\zeta), \, \nonumber \\
\frac{\partial \left<T\right>}{\partial z} &=& \frac{T_\ast}{\kappa \vert z \vert } \phi_h(\zeta), \, \nonumber \\
P &=& \frac{u_\ast^3}{\kappa \vert z \vert} \phi_m(\zeta), \, 
\end{eqnarray}
where $T_\ast = Q_{net}/(\rho_0 c_p u_\ast$). Reviews of experimental data in the atmospheric boundary layer \citep{BusingerWIB:1971, Hogstrom:1988, Hogstrom:1996} have suggested the following forms for the universal stability functions: 
\begin{flalign}
\phi_m(\zeta) &= 1 + 5.3 \zeta, \quad \quad \phi_h(\zeta) = 1 + 8 \zeta \nonumber \\
& \text{for} \quad 0 < \zeta < 0.5, \, \nonumber \\
\phi_m(\zeta) &= (1 - 19\zeta)^{-1/4}, \quad \phi_h(\zeta) = 0.95 (1 - 11.6\zeta)^{-1/2}  \nonumber \\
& \text{for} \quad -2 < \zeta < 0.
\end{flalign}
It is noted that the pre-multiplication factors in front of $\zeta$ have uncertainties. Recent observational data in the Southern Oceans and LES of the convective surface mixed layer have also raised question of MOST validity \citep{LargePD:2019, ZhengHD:2021, GiddyWF:2026}.

 \begin{figure*}[t!]
	\noindent\includegraphics[width=39pc,angle=0] {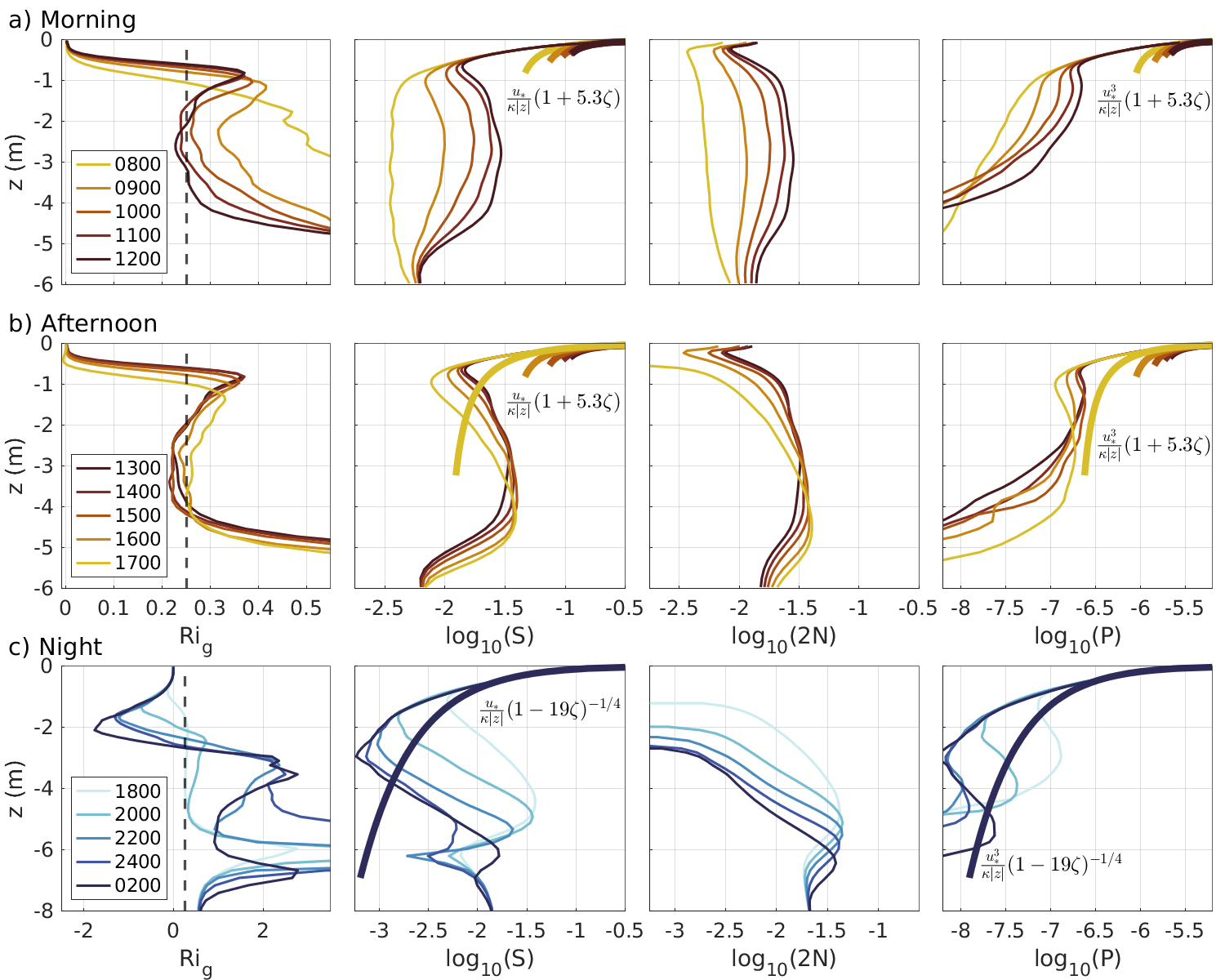}\\	
	\caption{Comparison between the present LES results and the MOST scaling in the surface mixed layer during: (a) morning; (b) afternoon; and (c) night of the first day of simulation in case L45H6C1. From left to right, each column shows the vertical profiles of $Ri_g$, shear $S$ (including Stokes contribution), stratification $N$, and turbulent production $P$. In the second and fourth columns, thick curves indicate the scalings of MOST. Note that MOST scalings are only shown in the regions where $0 < \zeta < 0.5$ for stable forcing conditions (panels a, b) and $-2 < \zeta < 0$ for unstable forcing conditions (panel c). The vertical dashed lines in the panels of the first column indicate the critical value, $Ri_g = 0.25$.}
	\label{fig:MI_MOST}
 \end{figure*}

The simulations in the present study imply that MOST is not applicable to the stable forcing conditions of the DWL as demonstrated in Fig.~\ref{fig:MI_MOST}. The morning evolution (Fig.~\ref{fig:MI_MOST}a) during the first day of case L45H6C1 is discussed first.
As the MI develops during the morning hours, the $Ri_g$ profiles in the first column of Fig.~\ref{fig:MI_MOST}(a) shows the thickening of MI layers between 0800 and 1200 hrs with values clustering around 0.25. The shear and stratification grow consistently by almost an order of magnitude during this time period (second and third columns of Fig.~\ref{fig:MI_MOST}a) while maintaining the near-critical values of $Ri_g$. With the 
increasing shear, the turbulent production between 3 and 5-m depth increases by nearly a factor of three (fourth column and first row) at 1200 relative to 0800 hrs. 
The increase in shear and production occurs mostly outside the MOST layer that is limited to $\zeta < 0.5$ ($z > -0.9$ m).
The LES profiles show good agreement with the MOST scalings only in the thin region $\zeta < 0.1$ ($z > -0.2$ m), very near the surface.
Between 0800 and 1200, the shear profiles remain similar above 0.5-m depth, which is contrast with MOST.
According to the MOST stability function, $\phi_m = 1 + 5.3 \zeta$, the shear should increase from 0800 to 1200 at a given $z$ since the MO length ($L$) continually decreases and the value of $\zeta = \vert z \vert /L$ correspondingly increases.
The LES shear and turbulent production indicate the TKE above 0.5-m depth is not suppressed despite the increase in stratification.

 Previous LES studies of unstable surface forcing have suggested that Langmuir circulations can reduce the near surface shear when $La < 0.3$ \citep{LargePD:2019, ZhengHD:2021}. In the present study, we observe the reduction in shear (relative to MOST scaling) with the value of $La = 0.3$ and under stable forcing conditions. At 1200 hrs, the shear values at $\zeta = 0.5$ ($z = -0.35$ m) differ by nearly an order of magnitude between the LES and MOST predictions.
 Beyond numerical differences, the profiles also exhibit qualitatively distinct shapes. While the MOST profiles indicate monotonic decrease in shear and production with depth, the LES profiles show increasing values in the region $-0.8 > z > -1.1$ m.

MI and the deviation from MOST are also observed during the afternoon period (Fig.~\ref{fig:MI_MOST}b). 
Between 1300 and 1700, while the $Ri_g$ values between 1.5 and 4-m depths cluster in a narrow band around 0.25 and the stratification decreases due to mixing (first and third columns), the shear and production profiles remain clearly different from the MOST scalings (second and fourth columns). 
The MOST shear value at the depth of $\zeta = 0.5$ ($z = -3.3$ m) at 1700 hrs is nearly three times smaller than the LES value. Nevertheless, the MOST production ($P$) is approximately {\em twice larger}. MOST assumes that the production is induced by wind-driven boundary layer turbulence (i.e. $u_*$), albeit modified by a stability function. The afternoon turbulence is driven by MI and the descending shear layer - a process that is qualitatively different from a canonical MOST-modified boundary layer. The LES profile of $P$ at 1700 hrs clearly shows the delineation between boundary layer turbulence and MI turbulence at z = -1 m. Specifically, with increasing depth, $P$ follows the MOST-based expectation of a downward trend only until z = -1 m, before it reverses and begins to increase.

The deviation from MOST continues on into the night between 1800 and 0200 hrs, when unstable forcing prevails (Fig.~\ref{fig:MI_MOST}c). 
The $Ri_g$ profiles deviate from the state of MI (left column) with negative values above 2-m depth and values considerably larger than 0.25 in the region below.
The MOST shear prediction (one profile that holds throughout the night) significantly underestimates the LES values at nearly all depths. They only agree in the very near surface layer $\zeta > -0.1$ ($z > -0.35$ m), outside of which the MOST shear can be as much as 30 times smaller.
It is interesting that the remnant of day-time DWL and MI can have great influence on the night-time evolution of ML turbulence.
For example, the LES production at 4-m depth is considerably larger than the MOST prediction between 1800 and 2000 hrs.
While LES indicates the prevalence of shear production in the night-time mixing layer, MOST incorrectly parameterizes the turbulence during the early night as convection.
Later in the night, there is significant shear but the TKE production is much lower than the MOST prediction.
These discrepancies put into question whether the MO-based stability correction over $-2 < \zeta < 0$ is applicable to the night-time turbulence found here.

 \begin{figure*}[t!]
	\noindent\includegraphics[width=39pc,angle=0] {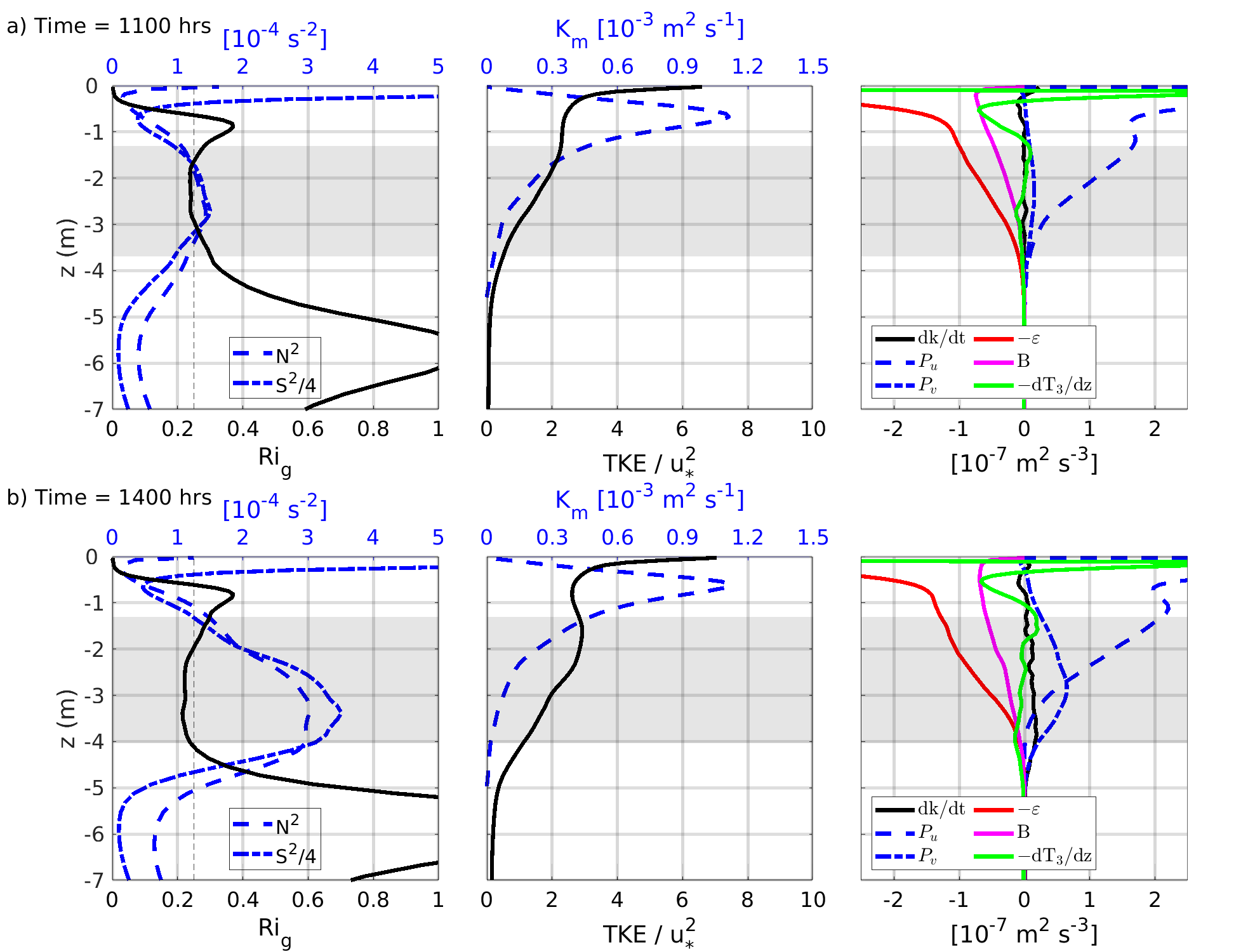}\\	
	\caption{Turbulent kinetic energy budget in the MI layer at (a) t = 1100 and (b) 1400 hrs on the second day in case L45H6C1. Profiles of $N^2$, $S^2$ and $Ri_g$ are shown in the left panel; eddy viscosity ($K_m$) and turbulent kinetic energy (TKE) in the middle panel; and TKE budget terms in the right panel. The budget terms include along-wind ($P_u$) and cross-wind ($P_v$) shear production of TKE, turbulent dissipation rate ($\varepsilon$), buoyancy flux ($B$), transport ($dT_3/dz$) and tendency ($dk/dt$). Vertical dashed lines in the left panel indicate $Ri_g = 0.25$. Shaded regions mark the MI layer.}
	\label{fig:MI_tkebud}
 \end{figure*}

\subsection{Turbulent kinetic energy budget}

To further understand the role of MI on the MOST deviation, we examine the TKE budget as illustrated in figure~\ref{fig:MI_tkebud} at two times: during the growth of MI turbulence at t = 1100 hrs and during its decay at t = 1400 hrs. Elevated $N^2$ and $S^2$, and the clustering of $Ri_g$ around 0.25 occur in the MI layer (shown as the grey shaded region) throughout the morning as is illustrated at 1100 hrs (Fig.~\ref{fig:MI_tkebud}a, left panel) while the TKE, shear production (including along-wind component, $P_u$, and cross-wind component, $P_v$) and turbulent dissipation rate ($\varepsilon$) indicate significant turbulence in the layer (Fig.~\ref{fig:MI_tkebud}a, middle and right panels). The significant value of negative $B$ indicates that MI turbulence actively mixes the buoyancy field in MI layer despite the stabilizing heat flux. 
It is noteworthy that the direct balance between $P_u$, $\varepsilon$ and $B$ that is exhibited at 1100 hrs is a feature that persists throughout the morning when background $N^2$ and $S^2$ increase consistently. 
Downward momentum transport supports the continuous enhancement of $S^2$ as indicated by the eddy viscosity profile ($K_m$, middle panel in Fig.~\ref{fig:MI_tkebud}a). 
 
By t = 1400 hrs in the afternoon, the peak values of $N^2$ and $S^2$ in the MI layer have increased approximately by a factor of 2 (Fig.~\ref{fig:MI_tkebud}b, left panel) as the MI layer enters a stage where shear and stratification continuously decay. Relative to the earlier time, the TKE and budget terms become larger and, notably, $P_v$ owing to the Ekman veer of the current becomes appreciable in the MI layer (Fig.~\ref{fig:MI_tkebud}b, middle and right panels). The balance between $P_u$, $B$ and $\varepsilon$ during the growth stage of mean shear and stratification is no longer observed as the tendency, transport and lateral shear production terms also contribute to the TKE budget.

\section{Variabilities in DWL and MI from the Equatorial to the polar oceans} \label{sec:lat}

 In this section, we analyze four cases - L15H6C1, L30H6C1, L45H6C1, and L70H6C1 - to demonstrate how MI develops and modulates ML turbulence across various latitudes. These simulations are forced with same wind stress ($\tau_w = 0.02$ N m$^{-2}$) and diurnal heat fluxes ($Q_{sw,max} = 600$ W m$^{-2}$ and $Q_{ns} = -100$ W m$^{-2}$). We will deduce that marginal instability is a defining characteristic of the DWL; it can manifest at all latitudes in the global oceans when the conditions of this study, namely, low wind and substantial solar heat flux apply. Furthermore latitude-dependent changes in the inertial response to the applied forcing lead to significant differences in the turbulent boundary dynamics, despite the prevalence of MI in all simulations.
 
 \begin{figure*}[t]
 	\noindent\includegraphics[width=39pc,angle=0] {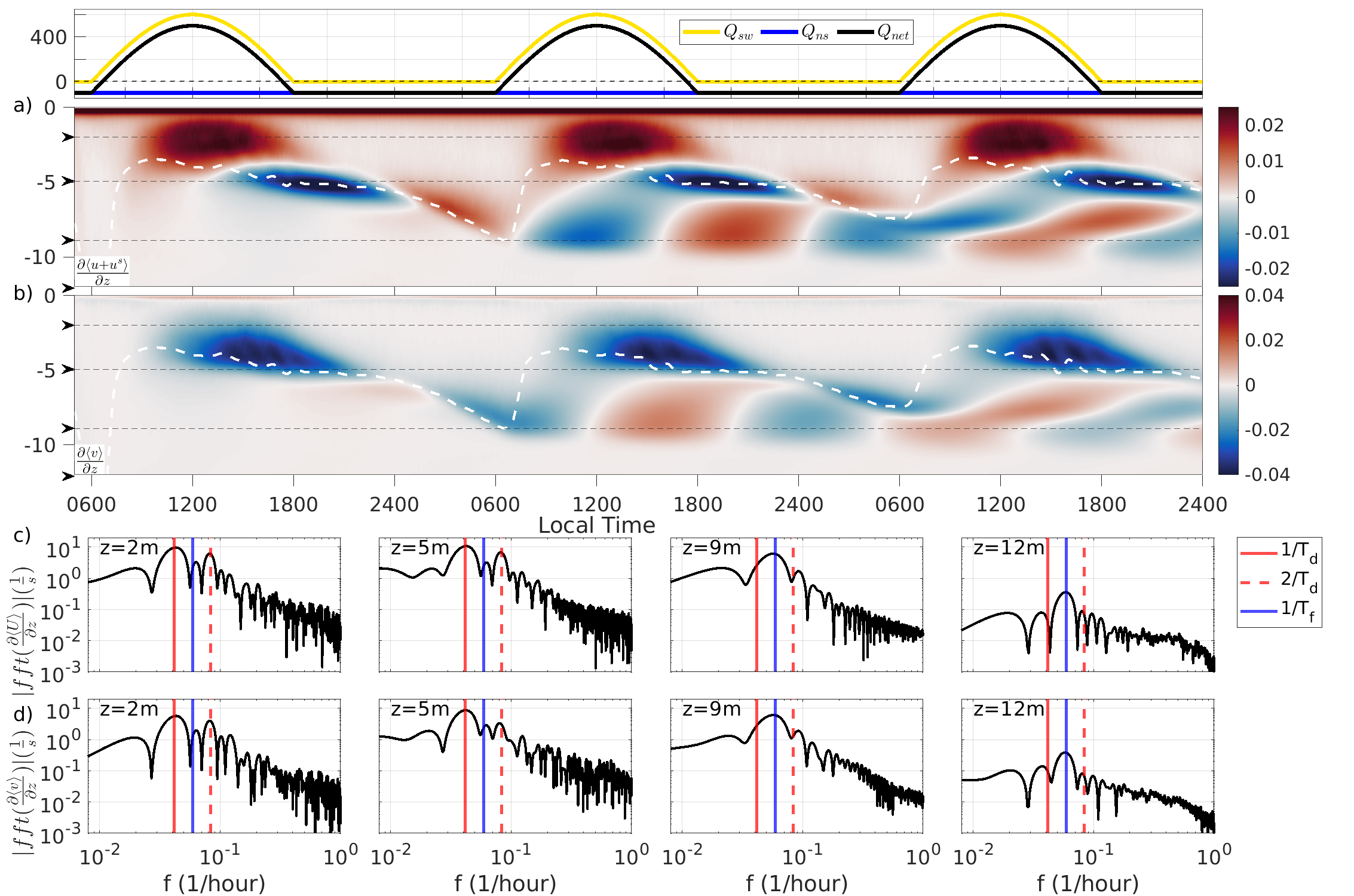}\\
 	\caption{Vertical shear of (a) zonal velocity ($\left<u+u^s\right>$) and (b) meridional velocity ($\left<v\right>$). Both panels show modulation at the diurnal time scale ($T_d$), the solar heating period ($T_d/2$), and also the inertial time scale ($T_f$) in case L45H6C1. Here $T_d$ = 24 hrs and the inertial period $T_f$ = 17 hrs. Panels (c, d) shows the frequency spectra of zonal and meridional shear, respectively, at various depths. The spectra at 9-m and 12-m depths show the strongest peak at the inertial frequency ($1/T_f$), while the spectra at 2-m and 5-m depths exhibit the strongest peak at the diurnal frequency ($1/T_d$).}	
 	\label{fig:dUdz_FFT}
 \end{figure*}

The time history and temporal spectra of Fig. \ref{fig:dUdz_FFT} illustrate the interaction between the inertial shear and diurnal shear in the L45 simulation. The vertical gradient of zonal velocity (Fig. \ref{fig:dUdz_FFT}a) exhibits diurnal oscillations with a period of $T_d$ = 24 hrs in the top 3-m layer showing positive values during the day and zero nightly shear due to convective mixing. 
The bands of negative zonal shear associated with the descending shear layers occur daily between 3- and 6-m depth in the late afternoon. 
The inertial shear manifests in the deeper region between 6- and 10-m depth as alternating bands of positive and negative shear with a period of $T_f$ = 17 hrs. The amplitude of the inertial shear bands decreases sharply below 10-m depth.
The DWL formed on the first day is unaffected by inertial shear, as there is little motion below the ML. However, the interaction between diurnal and inertial shears is evident during the second night when the MLD at 0600 hours is notably shallower than the corresponding value after the first night. Rotation acts to reduce turbulent entrainment as previously shown in the studies of Ekman currents \citep{PollardRT:1973, PhamS:2017}.

 Zonal shear spectra at various depths between 2 and 12 m in Fig. \ref{fig:dUdz_FFT}(c) highlight the relative contribution of inertial shear to boundary layer dynamics. The spectra at 2- and 5-m depth shows the two strongest peaks at the time periods of 12 and 24 hours (red dashed and solid line, respectively) corresponding to the diurnal cycle of the heat flux. The third strongest peak corresponds to the inertial time period of 16.9 hrs (blue solid lines). In contrast, the spectra at 9- and 12-m depth below the MLD exhibit the strongest peak at the inertial period and no distinct peaks at the 12- or 24-hr time periods. In this particular L45 simulation, the inertial and diurnal shears evolve out of phase of each other. The negative and positive inertial shear bands that develop on the second day have little effect on the MLD; however, the following days's negative band correlates with the shallowing MLD when comparing with the first night. 

 \begin{figure*}[t]
	\noindent\includegraphics[width=39pc,angle=0] {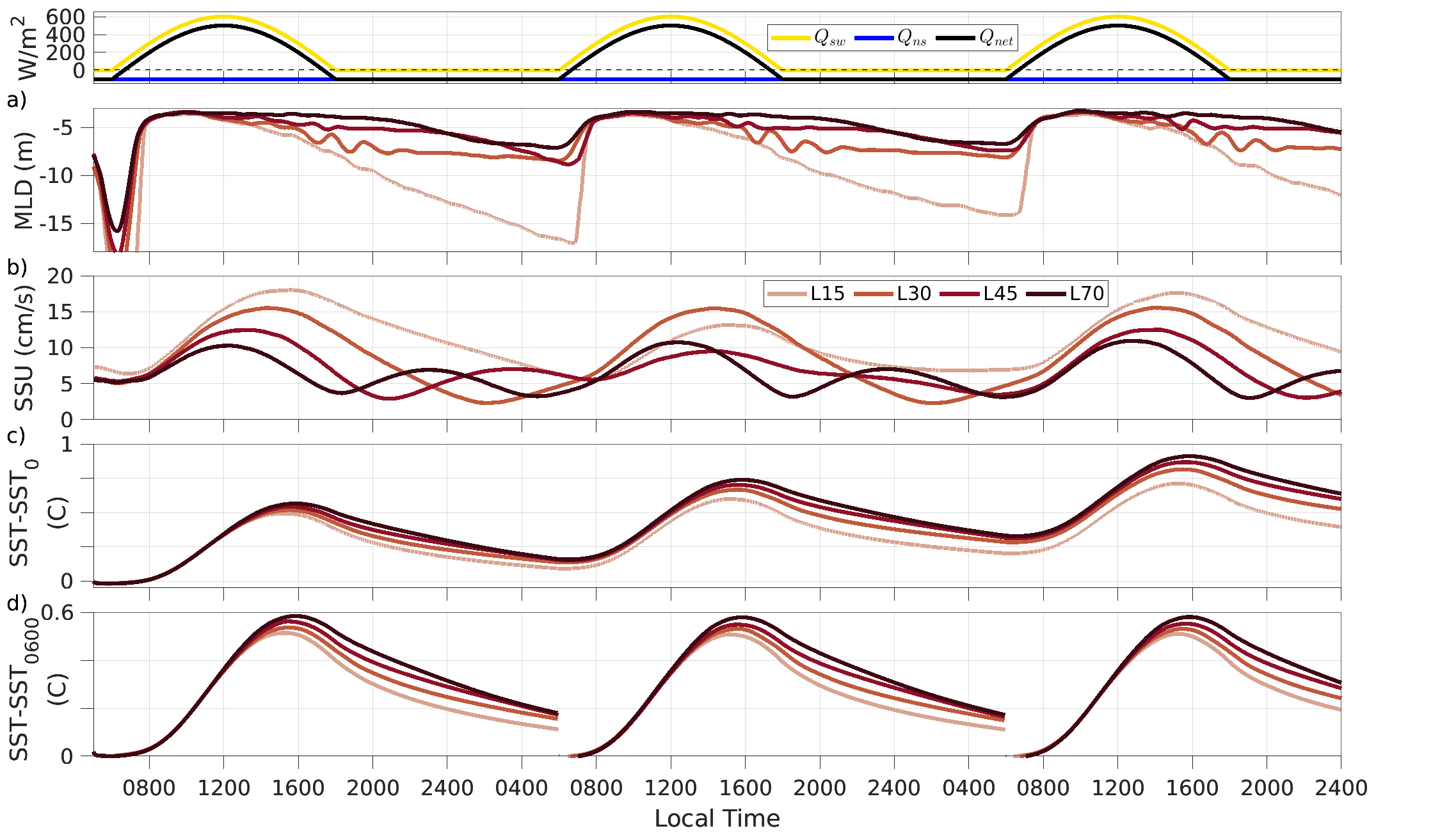}\\
	\caption{Effect of rotation on the near-surface boundary layer at different latitudes in cases L15H6C1, L30H6C1, L45H6C1, and L70H6C1: (a) MLD; (b) sea surface speed, SSU; (c, d) sea surface temperature, SST, and its deviation from the value at 0600 hrs, $SST - SST_{0600}$. Under the same wind stress and surface heat fluxes, the daily peak SST increases at higher latitudes while the peak value of SSU and entrainment rate decrease.} 
	\label{fig:Lat_surf}
 \end{figure*}

 The effects of rotation on the MLD, sea surface velocity magnitude (SSU including the Stokes drift) and temperature (SST) are illustrated in Fig.~\ref{fig:Lat_surf}. The minimum MLD (Fig. \ref{fig:Lat_surf}a) occurs in the morning, and is similar regardless of latitude since the diurnal solar heat flux (invariant among cases) controls the shallowest MLD values. 
 However, the subsequent evolution of MLD varies significantly with latitude.
 As the DWL evolves in the afternoon, the stronger Ekman veering at high latitudes slows down turbulent entrainment. The MLD in the L15 simulation grows rapidly during the afternoon while the L70 simulation shows little change. The MLD in the L30 and L45 simulations exhibit high-frequency fluctuations during the entrainment process; the MLD deepens in bursts with a time period of approximately 1.5 hr. 
 
 The nightly entrainment also differs among cases with different latitudes. During late night and early morning of the second simulated day, the shallower ML at higher latitudes entrains faster. The L70 simulation shows that the MLD grows from 4-m depth at 2000 hrs to 7-m depth at 0600 hrs of the following morning. In contrast, the MLD in the L30 simulation exhibits a more moderate growth - from 7.5 to 8.2-m depth - over the same period.
The exception to this trend is at 15$^\circ$N, where the low value of $f$ has little effect and, therefore, the entrainment rate is largely unchanged from afternoon to overnight.
Increasing the relative strength of rotation (i.e., shortening the inertial period) reduces the shear-driven entrainment rate of the descending shear layer in the afternoon but enhances the entrainment rate at late night.
The late-night fast entrainment at high latitude is driven by the night-time enhancement of inertial shear. The SSU (Fig. \ref{fig:Lat_surf}b) show a second peak during the night in the L45 ($T_f$ = 16.9 hrs) and L70 ($T_f$ = 12.8 hrs) simulations in addition to the first peak that occurs at daytime. The second peak of SSU manifests as enhanced inertial shear that induces the late-night rapid entrainment.

Overall, the SSU (Fig. \ref{fig:Lat_surf}b) shows strong modulation at both inertial and diurnal time scales. Over the simulated three-day period, Ekman veering lowers the SSU as $f$ increases, leading to a 40\% lower peak SSU at 70$^\circ$N when compared to 15$^\circ$N on the first day. The relative phase of the inertial and diurnal signals varies in this three-day realization and depends on initialization.
At 45$^\circ$N, the inertial oscillation develops out of phase with the diurnal heating flux on the second day, leading to a reduction in peak SSU when compared to the peak SSU on the first day. On the third day, when the two modulations are closer to being in phase, the SSU rebounds to the same peak value as on the first day. 
 
Figure~\ref{fig:Lat_surf}(c) contrasts the SST variability among low and high latitudes. All simulations display a similar trend of morning SST warming until 1400 hrs, after which the SST starts to plateau for a few hours and then cool in the night. Overall, the SST increases over the three-day period due to the net downward surface heat flux in all cases.
The daily peak SST values and the night-time cooling rates depend on latitude. At low latitudes (e.g., 15$^\circ$N and 30$^\circ$N), the deeper MLD induced by the faster entrainment rate of the descending shear layer in the afternoon results in the faster cooling of SST. 
To isolate of the effect of the DWL on the daily SST warming, we plot the SST deviation from the 0600-hr values for each of the three days in Fig. \ref{fig:Lat_surf}(d). At higher latitude, the SST exhibits larger peak values which occur later in the afternoon. The SST increase is largest in the L70 case with the warming of 0.6$^\circ$C occurring at 1600 hrs (4 hours after the peak solar heat flux). In contrast, the L15 case shows a peak SST deviation of 0.5$^\circ$C that occurs at approximately 1500 hrs. For a more direct comparison of DWL characteristics, the duration of solar insolation has been kept constant at all latitudes. At high latitudes, summer conditions with longer periods of solar heating than nighttime cooling will lead to even higher SST warming.

 \begin{figure*}[t!]
	\noindent\includegraphics[width=39pc,angle=0] {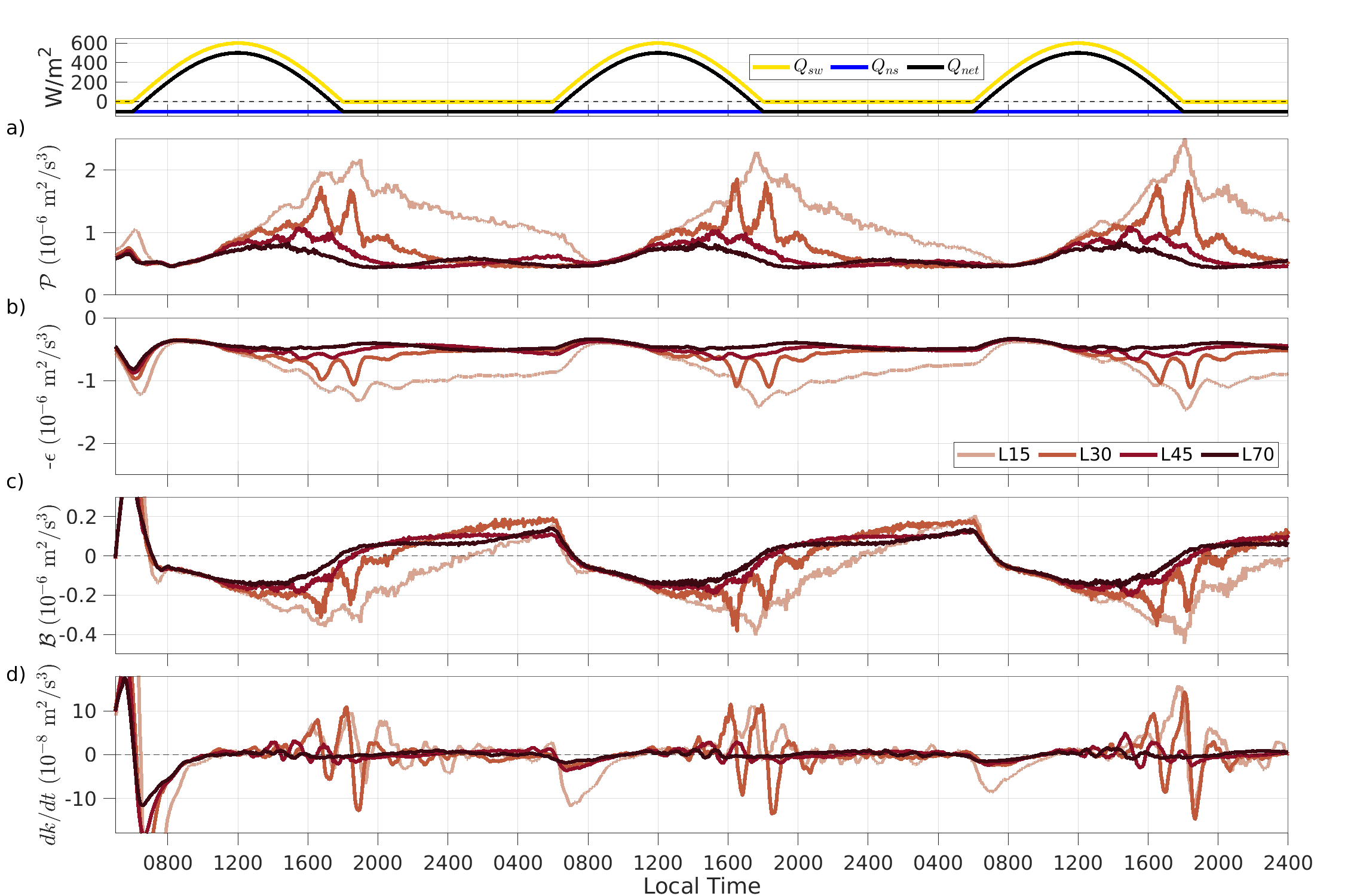}\\
	\caption{ The latitudinal dependence exhibited by terms in the depth-integrated TKE transport equation: a) Production, $P$; (b) dissipation, $\varepsilon$; (c) buoyancy flux, $B$; and (d) TKE tendency, $dk/dt$. The wind stress and surface heat flux are held constant among these cases. The peak values of the TKE budget terms decrease with increasing latitude in cases L15H6C1, L30H6C1, L45H6C1, and L70H6C1. Bursts of shear instability with a 1.5-hour period are prominent during the afternoon in the L15H6C1, L30H6C1 and L45H6C1 cases. Differences in the first 3 hours (until 0900 hrs) of the time history are due to different initial conditions after the spin-up periods.} 	
	\label{fig:Lat_TKE}
 \end{figure*}

Stronger turbulent entrainment typically signifies stronger TKE production and dissipation rate. Fig.~\ref{fig:Lat_TKE} (a,b) indicates that the depth-integrated values of TKE production ($\cal{P}$) and dissipation rate ($\cal{\epsilon}$) decrease with increasing latitude. 
The peak $\cal{P}$ in case L15 is approximately 2.5 times larger than in case L70. Peak $\cal{P}$ occurs at approximately 1500 hrs in L70 while it occurs significantly later (approximately 1900 hrs) in L15. 
Case L30 exhibits distinct bursts of production and dissipation with a period of $\sim$1.5 hours similar to the bursts in MLD previously shown in Fig.~\ref{fig:Lat_surf}(a). 
The bursts are also observed in the L15 and L45 simulations, however, with smaller amplitudes; and no burst is observed in L70.
The effect of inertial oscillations on the SSU (recall Fig. \ref{fig:Lat_surf}b) is also observed in $\cal{P}$. At approximately one inertial period after the solar heating peaks in L45 and L70, $\cal{P}$ exhibits a second peak, late in the night. 

The depth-integrated buoyancy fluxes ($\cal{B}$) shown in Fig.~\ref{fig:Lat_TKE}(c) further demonstrate the effect of latitude. Over a 24-hour period, the L15 simulation shows negative $\cal{B}$ for 18 hours while $\cal{B}$ takes positive values for only 6 hours between 2400 to 0600 hours. This suggests the dominance of shear-driven turbulence in the surface boundary layer that drives mixing of the buoyancy field, even when the surface heat flux indicates cooling between 1800 to 2400 hrs. In contrast, $\cal{B}$ changes sign from negative to positive values between 1800 and 2000 hours at other latitudes. In tropical oceans, shear-driven turbulence dominates the surface boundary layer even during night-time convective forcing. At mid - and high latitudes, the contribution of night-time convection becomes relatively more dominant, noting that the same surface flux was applied for all four cases. The TKE tendency ($dK/dt$) in Fig.~\ref{fig:Lat_TKE}(d) re-emphasizes the 1.5-hour bursts of shear instabilities / MI in the afternoon of the L15 and L30 cases, corresponding to the previously-shown high-frequency fluctuations in MLD, shear and production.

 \begin{figure*}[t!]
	\noindent\includegraphics[width=39pc,angle=0] {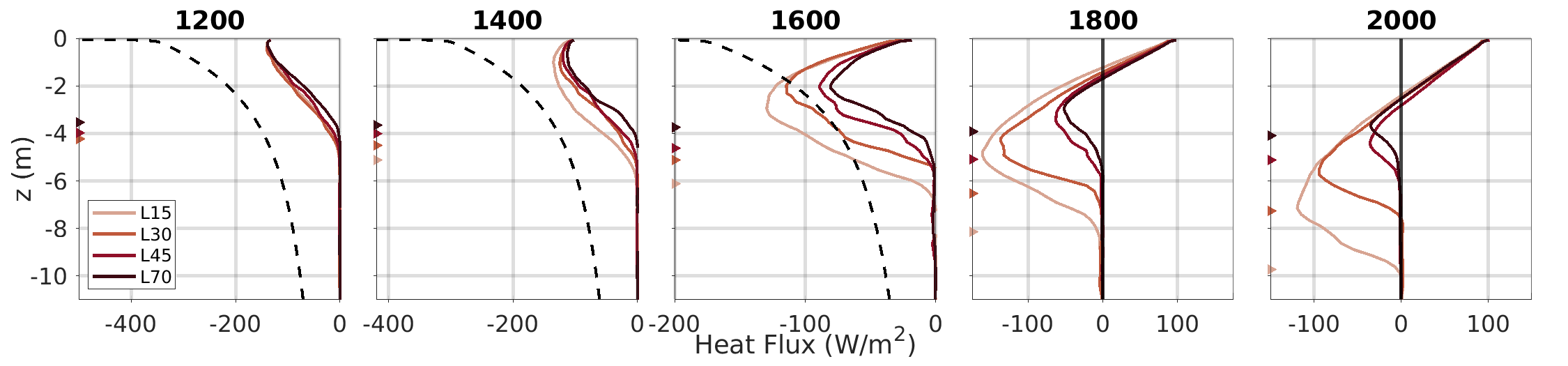}\\
	\caption{Relative strength between the penetrative solar radiation ($ - Q_{sw}$ shown with dashed lines) and turbulent heat flux ($J_q$, solid lines) at various instants during the second simulated day in cases L15H6C1, L30H6C1, L45H6C1, and L70H6C1. As latitude increases with the same surface heat flux and wind, the peak $J_q$ decreases in magnitude and is at a depth closer to the MLD (marked by similarly colored triangles).}
	\label{fig:J_Q}
 \end{figure*}

 \begin{figure*}[t!]
	\noindent\includegraphics[width=39pc,angle=0] {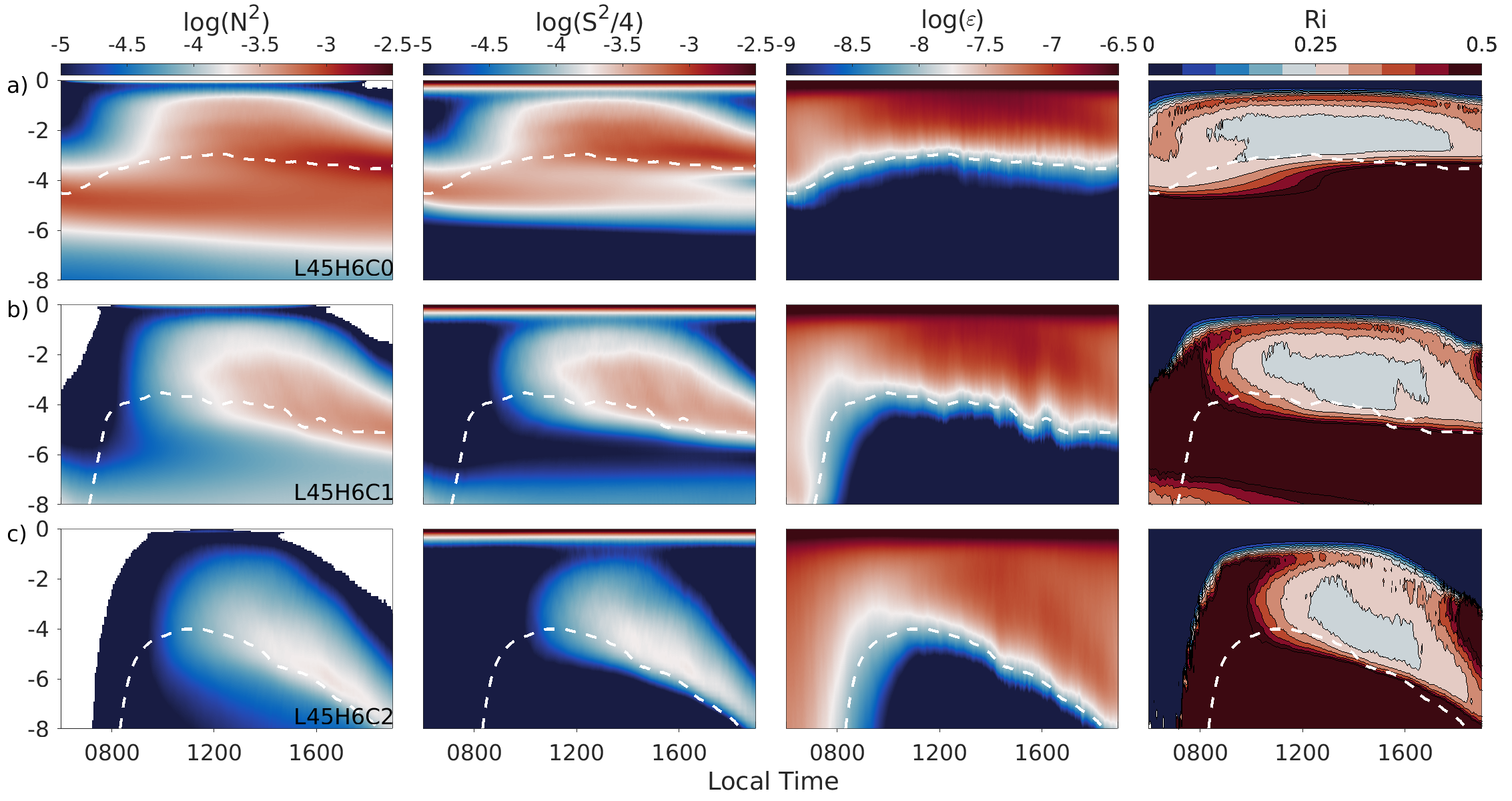}\\
	\caption{Increasing convective heat flux (i.e, decreasing $Q_{net}$) while maintaining the same wind, $Q_{solar}$ at 45$^\circ$N latitude in cases L45H6C0, L45H6C1 and L45H6C2 decreases stratification (first column), reduced Lagrangian shear (second column), and peak dissipation (third column) in the DWL The stratification and shear are reduced proportionally, so that the state of marginal instability with $Ri_g \approx 0.25$ (last column) is sustained over a wide range of convective forcing.}	
	\label{fig:Cool_NSE}
\end{figure*}

{
Figure~\ref{fig:J_Q} contrasts the turbulent heat flux, $J_q = \rho_0 c_p \left(\langle T'w'\rangle - \kappa_T\frac{\partial \langle T \rangle}{\partial z}\right) + Q^{T,sgs}_3 $, among the four cases at various times during the second simulated day. 
According the Reynolds-averaged temperature equation, 
\begin{equation*}
	\frac{\partial \langle T \rangle}{\partial t} =   - \frac{1}{\rho_0 c_p}  \frac{\partial} {\partial z}  \left( J_q - Q_{sw}\right), \, 
\end{equation*}
the positive (negative) $J_q$ values indicate upward (downward) heat flux, and its divergence correlates with the rate of warming. We note the minus sign in front of $Q_{sw}$ (a positive quantity) indicates downward solar radiation in a coordinate system with $z$ pointing up.}
The penetrative solar flux tends to warm the ML as can be seen mathematically by the positive contribution of the divergence of $Q_{sw}$ to $\partial \langle T \rangle/\partial t$ in the temperature transport equation. At noon, the mixing layer is warmed mainly by solar penetration since the divergence of $Q_{sw}$ is significantly larger than the divergence of $J_q$. 

At 1400 hrs, $J_q$ in the mixing layer becomes stronger with larger magnitudes occurring in the cases with lower latitudes. Noting that $J_q$ peaks in the middle of the ML, the divergence of $J_q$ has opposite sign in the upper and lower halves of the ML. 
The positive $\partial J_q/ \partial z$ in the upper half of the ML, which contributes to the cooling of local temperature, tends to counteract the warming by the penetrative solar flux. In contrast, the negative $\partial J_q/\partial z$ in the lower half of the ML tends to warm that portion through turbulent transport.
 At L15, the peak $J_q$ at 2-m depth reaches up to 70\% of $Q_{sw}$ at the same depth, signifying the increasing importance of turbulent thermal transport by MI at this time. ML turbulence continues to grow in the afternoon while the solar flux decreases. Consequently, the peak value of $J_q$ at 1600 hrs exceeds the $Q_{sw}$ value at the same depth in the L15 and L30 simulations. 
 Furthermore, the positive $\partial J_q/ \partial z$ in the upper half cools the near-surface water faster than the warming due to downward solar heat flux, and thus, SST becomes cooler. Since the entrainment rate of the descending shear layer is stronger at lower latitudes, the peak $J_q$ and its divergence are also larger leading to stronger cooling of SST at lower latitudes.

{When the surface heat flux is convective (positive $J_q$ values near the surface) between 1800 and 2000 hrs, the negative $J_q$ near the bottom of the ML shows stronger amplitude at lower latitudes leading to stronger positive $\partial J_q/\partial z$, and thus, the continuous stronger cooling of SST (Fig.~\ref{fig:Lat_surf}d).}
Near the MLD in the late afternoon and evening, heating from $\partial J_q / \partial z$ is 50\% stronger at L15 relative to L70. Within the top three meters, the difference in $\partial J_q/\partial z$ between latitudes decreases as the night goes on. 

 \begin{figure*}[t]
	\noindent\includegraphics[width=39pc,angle=0] {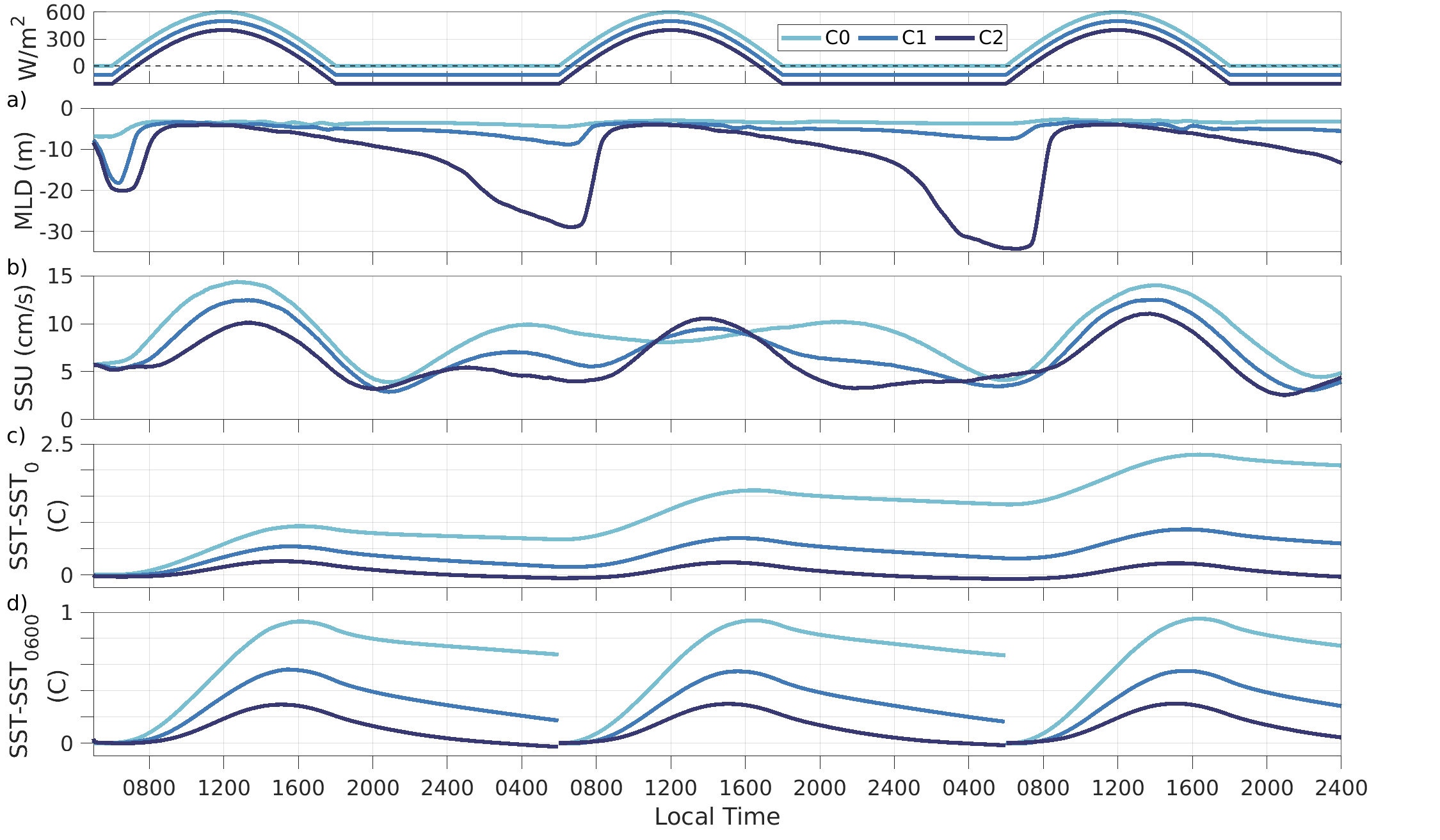}\\
	\caption{Increasing convection, and thereby decreasing $Q_{net}$, in cases L45H6C0, L45H6C1 and L45H6C2 results in: (a) deeper MLD; (b) slower surface jet SSU; and (c, d) a lower peak SST and its deviation from 0600-hr value, SST-SST$_{0600}$.}
	\label{fig:Cool_surf}
\end{figure*}

\section{Impact of changing net surface heat flux} \label{sec:convection}

In this section, we examine the effects of net surface heat flux ($Q_{net}$) that is changed by increasing the level of cooling flux in the three simulations L45H6C0, L45H6C1, L45H6C2, keeping other parameters constant. As $Q_{net}$ increases, the near-surface stratification increases; however, the shear increases in proportion such that the MI state is sustained in the DWL for all three cases. The turbulence varies significantly with $Q_{net}$ despite the same narrow range of $Ri_g$ realized in all cases. 

Figure~\ref{fig:Cool_NSE} compares the evolution of $N^2$, $S^2$, $\varepsilon$ and $Ri_g$ among the three cases. As $Q_{net}$ decreases from C0 to C2 cases, $N^2$ and $S^2$ in the DWL weaken consistently leading to weaker $\varepsilon$ despite the similarly narrow range of $Ri_g$ values around $0.25$. The DWL exhibits the state of MI in all three cases. Increasing convection (weakening $Q_{net}$) delays the onset of MI, i.e., $Ri_g$ dips below the critical value of 0.25 at a later time in C2. Stronger convection also results in faster deepening of the descending shear layer (due to weaker $N^2$ in the DWL) during the afternoon despite the relatively weaker local $\varepsilon$ values in C2. 
 Since $N^2$ and $S^2$ are not known {\em a priori}, it is difficult to parameterize MI-induced mixing in terms of the applied wind stress and surface heat fluxes especially when the MI layer does not follow MOST scaling as previously shown in section~\ref{sec:MI}.

We now move to the effect of changing the surface flux on the evolution of MLD and surface variables. Increasing $Q_{ns}$ affects entrainment by both the descending shear layer in the afternoon and late-night entrainment with the latter being larger as illustrated by the MLD evolution in Fig.~ \ref{fig:Cool_surf}(a). 
There is little entrainment in C0 with the MLD varying by less than 1.25 m over the first 24-hr period. Case C1 shows relatively stronger entrainment; the ML deepens from 3.5 m at 1000 hrs to 9 m at 0600 hrs during the first simulated day. 
Increasing the cooling flux further in case C2 causes the MLD to extend down to 29.1-m depth, an increase by more than 3 times when compared to C1 over the same time period. 

 \begin{figure*}[t!]
	\noindent\includegraphics[width=39pc,angle=0] {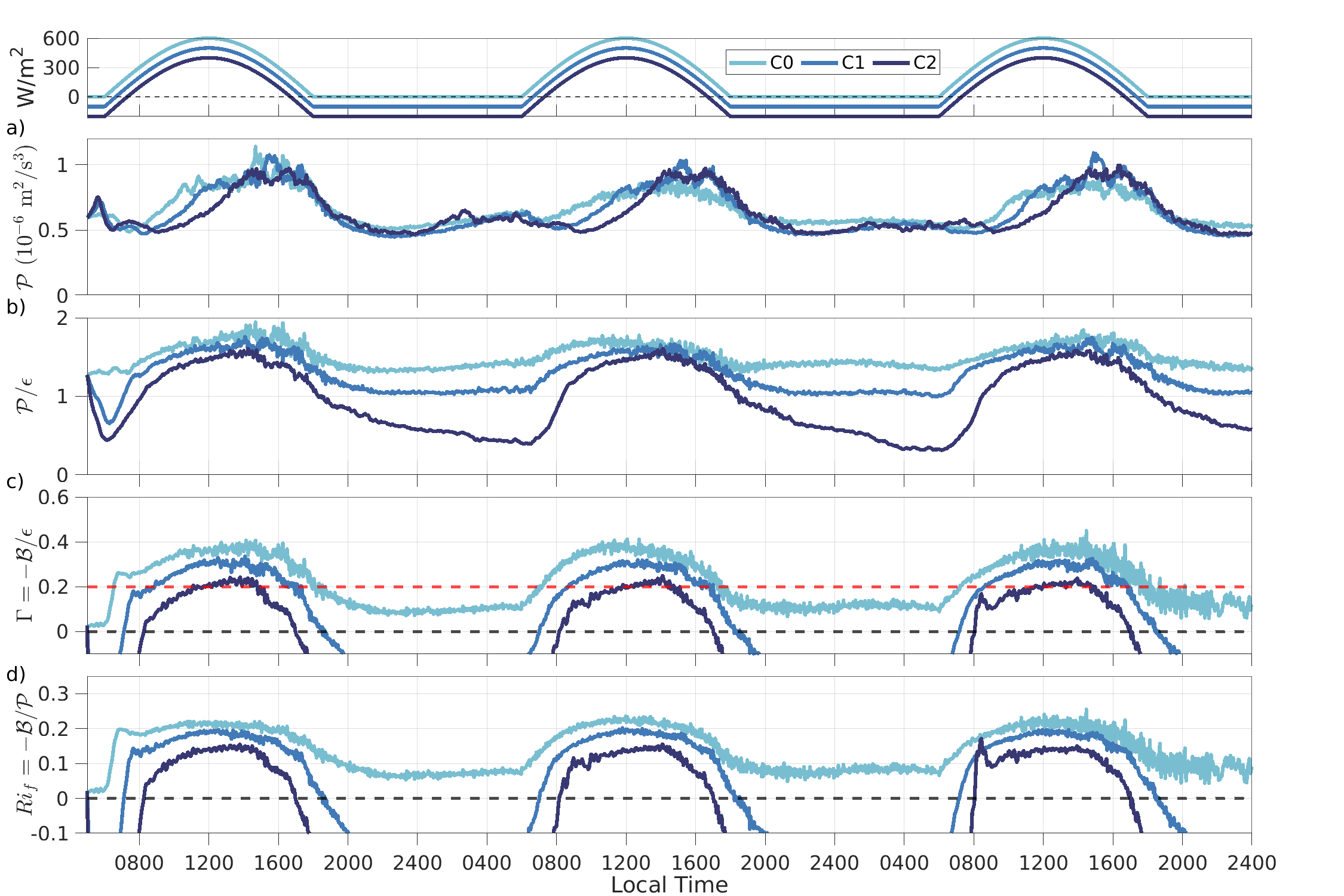}\\
	\caption{Effect of cooling flux on the depth-integrated TKE budgets: (a) Production, $\cal{P}$; (b) the ratio of integrated production to integrated dissipation, $\cal{P}/\cal{\varepsilon}$; (c) flux coefficient, $\Gamma = -\cal{B}/\cal{\varepsilon}$; and (d) flux Richardson number, $Ri_f = -\cal{B/P}$. The red dashed line in (c) denotes the value of 0.2 typically used in mixing parameterization \citep{Osborn:1980}.}
	\label{fig:Cool_TKE}
\end{figure*}

Notably, increasing the cooling flux while holding $Q_{sw}$ constant also affects the evolution of the surface jet (Fig.~\ref{fig:Cool_surf}b) with case-dependent differences of up to 5 cm$/$s. 
The SSU in case C0 exhibits inertial oscillations with four local peaks that are spaced at $T_f$ = 17.9 hrs apart. 
The first and last peaks have larger amplitude since they coincide with the timing of the peak solar heat flux leading to constructive interference that tends to accelerate the surface jet. The second and third peaks show weaker amplitude, and there is no peak at the time of maximum $Q_{net}$ during the second day. This implies destructive interference between the inertial oscillation and the diurnal solar flux at these times. If the inertial-oscillation phase with $U$ deceleration occurs around noon time, the strengthening of the surface jet by the solar heat flux is weakened resulting in the absence of a local peak in SSU. 
Increasing the cooling flux in C1 and C2 reduces the amplitude of the inertial oscillations and allows the acceleration of the SSU by the solar heat flux to be more prominent. Cases C1 and C2 show SSU peaks around noon on the second day.
Such inertial/diurnal interaction is a unique difference feature of the DWL at mid and high latitudes.
In equatorial oceans (i.e., absence of rotation effects), decreasing $Q_{net}$ at noon tends to weaken the surface jet. In contrast, decreasing $Q_{net}$ at mid latitude can strengthen the surface jet at noon due to the reduced effects of inertial oscillations. 

The evolution of SST shows larger variability in the cases with weaker cooling flux (Fig.~\ref{fig:Cool_surf}c,d). The SST increases by 0.93, 0.56 and 0.3$^\circ$C over the first morning period in the C0, C1 and C2 cases, respectively. A similar increasing trend is observed during the other two days. While the inertial oscillations influence the evolution of the diurnal surface jet from day to day, it does not affect the daily warming / cooling of SST. Recall that the SSU shows relatively stronger peaks around the noon time on the first and the last days and weaker peaks on the second day (Fig. \ref{fig:Cool_surf}b); however, there is no difference in the daily SST variability among the three days (Fig. \ref{fig:Cool_surf}d) for each case. 

With the same wind and solar forcing, increasing convective cooling has little impact on peak integrated shear production (Fig.~\ref{fig:Cool_TKE}a).
Increasing cooling flux delays the growth of turbulent production during morning time since the DWL takes longer to develop at smaller $Q_{net}$. 
 Recall the second column of Fig.~\ref{fig:Cool_NSE}, which shows that shear develops later in the day and decreases in amplitude as cooling is increased. 
 While the majority of the production occurs during the afternoon when the DWL becomes unstable to shear instabilities, there appears to be some production during the first night induced by the shear of the inertial oscillations.
 
  \begin{figure*}[t!]
 	\noindent
 	\begin{center}
 		\includegraphics[width=30pc,angle=0] {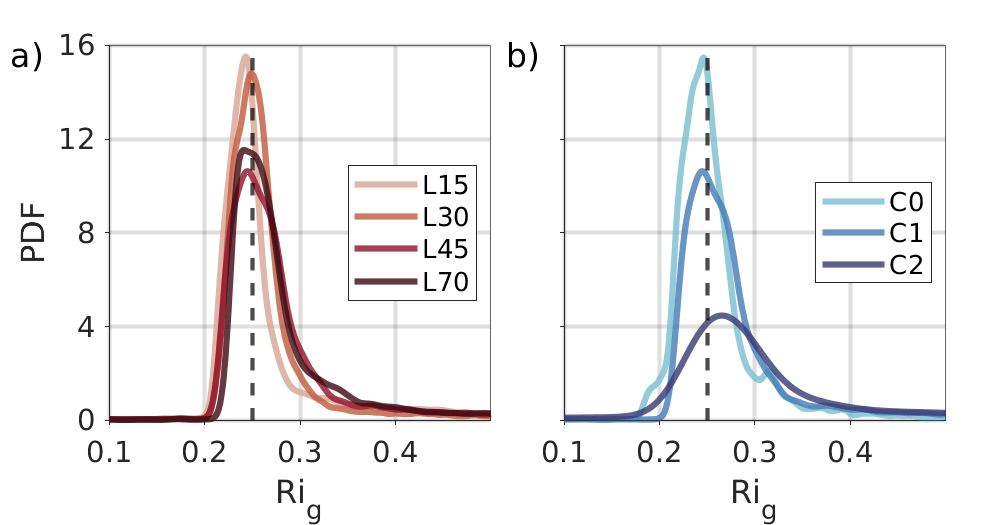}\\
 		\caption{Probability density functions of $Ri_g$ within the DWL and below the Stokes layer indicate the presence of marginal instabilities in all simulated cases. The mode of all cases remains between $Ri_g = 0.24$ and $0.28$.}
 		\label{fig:Rig_PDF}
 	\end{center}
 \end{figure*}
 
Depth-integrated turbulence quantities are also diagnosed. Increasing convection slightly decreases the ratio ($\cal{P}/\cal{\varepsilon}$) of integrated production to dissipation at noon time of the first day by approximately 10\%, from 1.69 in case C0 to 1.48 in C2 (Fig. \ref{fig:Cool_TKE}b). Overall, the ratio values fall in range with previous numerical studies of stratified shear instabilities. \cite{VanDinePS:2021} reported $\cal{P}/\cal{\varepsilon}$ to take values between 1.5 and 1.8 in their direct numerical simulations (DNS) of a canonical high-Reynolds number stratified shear layer. The agreement suggests that the turbulent mixing in a DWL should be viewed as local shear-driven mixing events rather than as an extension of boundary-layer turbulence, i.e., against the applicability of MOST. We note that the variability of $\cal{P}/\cal{\varepsilon}$ is small over the entire diurnal cycle of C0 since stratified shear turbulence dominates; however, nightly values become significantly smaller in cases C1 and C2 since it is convection that drives their night-time turbulence. 

To further compare the turbulence in the DWL with stratified shear turbulence, Fig.~\ref{fig:Cool_TKE}(c, d) shows the evolution of the flux coefficient, $\Gamma = -\cal{B}/\cal{\varepsilon}$, and the flux Richardson number, $Ri_f = -\cal{B}/\cal{P}$. 
The $\Gamma$ values vary between 0.2 and 0.38 while $Ri_f$ shows values between 0.1 and 0.22 during the afternoons when the DWL becomes unstable. 
The $\Gamma$ values are larger than 0.2, the value typically used for mixing parameterization \citep{Osborn:1980}. 
Previous DNS studies of stratified shear layers indicated $\Gamma$ can have larger-than-0.2 values, e.g., \cite{Mashayek:2017, GreggDRK:2018, LiuKS:2022, PhamS:2023efm}.

\section{Conclusions} \label{sec:conclusion}

Using LES, we have performed a parametric study to examine the impact of latitude and surface cooling on the evolution of diurnal warm layers (DWL) in the global oceans.
The LES model, which includes the effect of surface waves (i.e, Langmuir turbulence), is driven by a constant wind stress and a heat flux consisting of diurnal solar radiation and constant cooling flux. The latitude spans tropical (15$^\circ$N) to polar regions (70$^\circ$N) allowing the study to quantify the effects of rotation on DWL turbulence. 
We also examine the impact of the cooling flux on the diurnal variability of the ocean surface layer.
All simulations are performed for a 3-day period, which is sufficient to capture the interaction between inertial oscillations and diurnal variability.

For the chosen moderate-wind condition of 3.1 m$/$s leading to a wind stress of $\tau_w = 0.02$ N$/$m$^2$, a DWL, which is characterized by the day-time increase in SST and the formation of a surface jet, develops in all cases. 
The DWL is found to exhibit turbulence in a state of marginal instability (MI) throughout its evolution. The MI state is characterized by an intricate balance between the momentum input by wind, the stabilizing solar heat flux and especially mixing by turbulence such that the gradient Richardson number is maintained near the critical value of 0.25 for shear instabilities (Figs.~\ref{fig:DWL_background}, \ref{fig:Cool_NSE}, \ref{fig:Rig_PDF}). 
Below the Stokes layer and within the DWL layer, the probability density function of $Ri_g$ (Fig.~\ref{fig:Rig_PDF}) shows a narrow band with a strong peak near $Ri_g = 0.25$ at all latitudes and cooling fluxes. 
The MI state persists from morning until late afternoon and it triggers the descending shear layer which is responsible for significant turbulent thermal transport from the surface to the ocean interior (Fig.~\ref{fig:J_Q}). The enhanced local shear by wind and stratification by solar radiation in the DWL during the MI state cannot be predicted using the Monin-Obukhov Similarity Theory (MOST) of a turbulent boundary layer (Fig.~\ref{fig:MI_MOST}). The energy pathway from MI to turbulence is through bursts of Kelvin-Helmholtz-like shear instability (Fig.~\ref{fig:MI_tem}) and the mixing parameters such as the flux coefficient and flux Richardson number resemble characteristics of isolated local shear-driven turbulent events (Fig.~\ref{fig:Cool_TKE}). \cite{HughesMS:2021} did not find turbulence in a state of MI in the DWL that developed under a moderate wind speed of {4 m$/$s} and surface heating, but it is worth noting that, unlike the present LES, that study used 2D simulations, a constant background eddy viscosity, and did not include Langmuir turbulence.

The new findings of the present study suggest a pressing need for assessing whether the class of 1-D mixing models that is based on MOST can successfully parameterize DWL turbulence. We find that DWL supports a regime of stratified shear turbulence, in which local shear production is directly balanced by buoyancy flux and dissipation.
The dissipation rate in the MI layer varies by at least an order of magnitude while the Richardson number clusters in a narrow band around $Ri_g \approx 0.25$. Furthermore, the growing shear and stratification during the development of the MI layer requires correct parameterization of downward turbulent flux from the Stokes layer (i.e., correct $z$-dependent eddy viscosity, see \cite{Noh:1996} and \cite{NohGR:2009}). Misrepresentation of the turbulent flux (e.g., using a constant eddy viscosity) would alter the MI evolution.

Our case study indicates the presence of strong inertial oscillations (IOs) at mid and high latitudes. The interaction between the IOs and the diurnal surface jet leads to modulation in sea surface speed (SSU), MLD and sea surface temperature (SST), (Fig.~\ref{fig:Lat_surf}). The daily minimum MLD is unaffected by changing latitude as that depth is controlled by the applied wind stress and heat fluxes. However, increasing the value of latitude leads to weaker turbulent entrainment, shallower MLD, warmer SST and weaker SSU. The terms in the TKE budgets and the turbulent heat flux ($J_q$) also become weaker at higher latitudes (Figs.~\ref{fig:Lat_TKE}, \ref{fig:J_Q}).
With the same amount of wind work and net downward heat flux, the SST increases faster at high latitudes leading to stronger daily modulation of SST. 
The daily SST variability at $70^\circ$N is 20\% larger relative to the values at $15^\circ$N. Thus, our LES results suggest that SST-modulated air-sea interaction phenomena can be as important in polar oceans as in the equatorial oceans, despite the weaker solar radiation at high latitudes.

Increasing the convective heat flux weakens the surface jet and stratification in the DWL. Nonetheless, these quantities reduce simultaneously in a manner that maintains the presence of a MI layer between the Stokes layer and the MLD (Fig.~\ref{fig:Cool_NSE}). Within those depths, cooling dampens the peak of the PDF (Fig.~\ref{fig:Rig_PDF}b), but the mode remains in a similar range of $Ri_g = 0.24$ to $0.28$. Increasing night-time convection delays the onset of the MI layer in the morning and enhances the entrainment rate of the descending shear layer in the afternoon due to the weaker $N^2$ in the DWL. Stronger convection also reduces the amplitude of IOs at inertial timescale; however, the interaction between inertial and diurnal shear can accelerate the diurnal jet at shorter timescale within an inertial period. 

Finally, we found the existence of MI in DWL to be robust, given low wind speed and sufficiently strong solar heat flux. LES with the solar heat flux applied at the surface boundary (not penetrative radiation) also shows the presence of MI. Changing the 9-component penetrative solar radiation model of \cite{PaulsonS:1981} to the two-component model of \cite{PaulsonS:1977} affects the evolution of heat transport and SST, but it does not change the presence of MI. 
The value of Hoenikker number Ho = 0.56 at noon time is smaller than the range between 1 and 2 in which the suppression of Langmuir circulations by stabilizing buoyancy flux has been reported \citep{MinN:2004, PanL:2026}. 
Whether MI exists when the OSBL is forced with stronger solar radiation or weaker wind leading to large Ho requires future study. It is also of interest to assess how high wind and preconditioning of upper-ocean shear and stratification modify the diurnal variability of stratified shear turbulence that is found in the present work.

\clearpage

%

%
%
%
%
%

%
%
%
%


\clearpage

\acknowledgments
The authors gratefully acknowledge the support of the US Office of Naval Research: grant number 
N00014-23-1-2831.

%
%
\datastatement
 The LES data can be found in \cite{HallArch:2026}. The data and software used to generate the figures will be posted upon acceptance of the manuscript.

%






%



 \bibliographystyle{ametsocV6}
 \bibliography{references_sorted}

\end{document}